\documentclass[twocolumn]{aastex631}
\usepackage{natbib,times}

\shorttitle{Galaxy clusters in DESI}
\shortauthors{Wen and Han}

\begin{document}

\title{A catalog of 1.58 million clusters of galaxies identified from the DESI Legacy Imaging Surveys}


\author{Z. L. Wen}
\affil{National Astronomical Observatories, Chinese Academy of Sciences,
     A20 Datun Road, Chaoyang District, Beijing 100101, China. hjl@nao.cas.cn}
\affil{CAS Key Laboratory of FAST, NAOC, Chinese Academy of Sciences}

\author{J. L. Han}
\affil{National Astronomical Observatories, Chinese Academy of Sciences,
     A20 Datun Road, Chaoyang District, Beijing 100101, China. hjl@nao.cas.cn}
\affil{CAS Key Laboratory of FAST, NAOC, Chinese Academy of Sciences}
\affil{School of Astronomy and Space Sciences,
       University of Chinese Academy of Sciences, 
       Beijing 100049, China}


\begin{abstract}

Based on the DESI Legacy Imaging Surveys released data and available
spectroscopic redshifts, we identify 1.58 million clusters of galaxies
by searching for the overdensity of stellar mass distribution of
galaxies within redshift slices around pre-selected massive galaxies,
among which 877,806 clusters are found for the first time. The
identified clusters have an equivalent mass of $M_{500}\ge
0.47\times10^{14}~M_{\odot}$ with an uncertainty of 0.2 dex. The
redshift distribution of clusters extends to $z\sim1.5$, and 
338,841 clusters have spectroscopic redshifts. Our cluster sample
includes most of the rich optical clusters in previous catalogs,
more than 95\% massive Sunyaev-Zeldovich clusters and 90\% ROSAT
and eROSITA X-ray clusters. From the light distributions of member
galaxies, we derive the dynamical state parameters for 28,038
rich clusters and find no significant evolution of the dynamical state
with redshift. We find that the stellar mass of the brightest cluster
galaxies grows by a factor of 2 since $z=1$.
\end{abstract}


\section{Introduction}

As the largest virialized systems in the Universe, clusters of
galaxies contain hundreds to thousands of galaxies, hot intracluster 
gas, and dark matter. They are tracers of large-scale structures and
place powerful constraints on cosmological parameters
\citep{aem11,bcc+14,hhw16}.
The dense environments and deep gravitational wells make galaxy
clusters as an ideal laboratory for understanding galaxy
formation \citep{lxm+08,wtc+12}, emission of intracluster medium
\citep{rbn+02,fgg+12} and the nature of dark matter
\citep{mgc+04,bad+20}. 

Galaxy clusters can be identified as the over-density peaks of the
galaxy distribution in optical/infrared images \citep{abe58}, or
detected by the X-ray emission of hot intracluster medium \citep{cf76}
or via the Sunyaev-Zeldovich \citep[SZ,][]{sz72} signals in millimeter
bands. The optical photometric surveys provide a large number of
galaxy clusters. In the early years, about ten thousand nearby
clusters were recognized from the single-band images of the Palomar
Observatory Sky Survey \citep{zhw63, aco89, gdl+03, ldg+04}. When
multi-band data of the Sloan Digital Sky Survey (SDSS) are available,
more than 150,000 galaxy clusters have been identified at redshifts
$z<0.8$ based on the red sequence of cluster member galaxies
\citep{kma+07, hmk+10, rrb+14, ogu14} or photometric redshifts of
galaxies \citep{whl09,whl12, spd+11,wh15, bsp+18}. More clusters of
galaxies were found from the deeper data of Hyper Suprime-Cam Subaru
Strategic Program, Dark Energy Survey and Wide-field Infrared Survey
Explorer (WISE) in the redshift range up to $z\sim1.5$
\citep{oll+18,wh21,wh22,abd+21,tgb+24}.

Recently, the Dark Energy Spectroscopic Instrument (DESI) Legacy
Surveys\footnote{https://www.legacysurvey.org/} have been carried out
and released data in three optical bands ($g$-band, $r$-band, and
$z$-band\footnote{We use $z$ for redshift in this paper and $z_{\rm
  m}$ for the $z$-band magnitude to avoid confusion.}) with one
magnitude deeper than the SDSS \citep{dsl+19}. Incorporating the
public DECam data, the DESI Legacy Surveys cover a sky area of more
than 20,000 deg$^2$. Several large cluster catalogs have been
published based on previous DESI data release 9 (DR9)
\citep{yxh+21,zsx+22,ynd+23}. The latest DR10 data cover a larger sky
area with some overlapped areas in the DR9 and include new $i$-band
data from the NOIRLab Data Archive, an excellent database for
identifying a larger sample of clusters to higher redshifts.

\begin{figure*}
\centering \includegraphics[width = 0.7\textwidth]{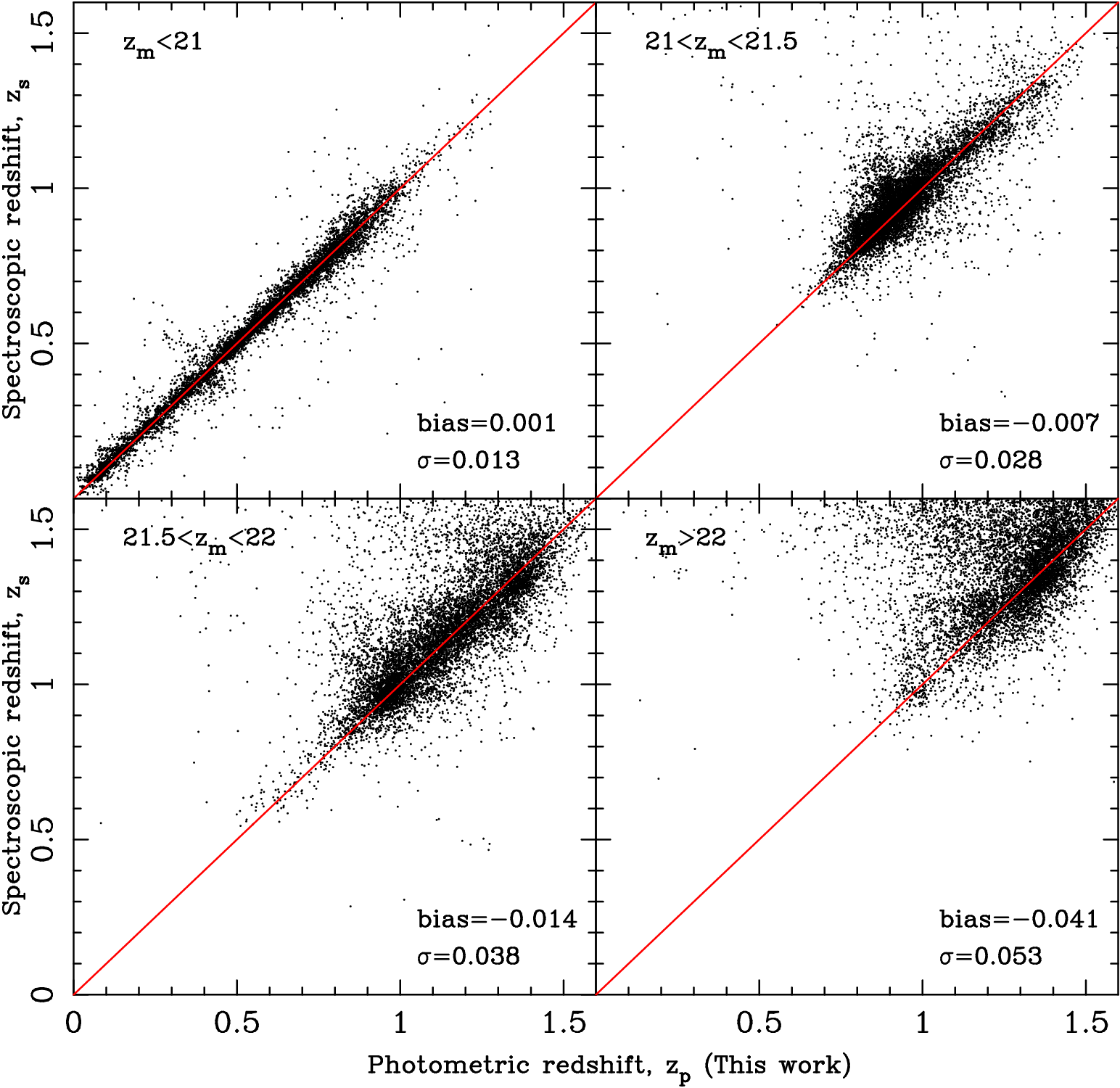}  
\caption{Comparison between the newly estimated photometric redshifts
  $z_{\rm p}$ and spectroscopic redshifts $z_{\rm s}$ for galaxies
    in the region of middle survey depth and in four $z$-band
    magnitude ranges with different uncertainty and systematic bias as
    marked in each panel, where only randomly selected 10,000 galaxies
    are taken for plotting.}
\label{photoztest}
\end{figure*}

Following our previous work \citep{whl09,whl12,wh15,wh21,wh22}, in
this paper we identify 1.58 million galaxy clusters from the DESI
Legacy Surveys released data by using photometric and available
spectroscopic redshifts of galaxies. We first estimate photometric
redshifts for galaxies with $griz$ data from the DESI Legacy Surveys
and mid-infrared data from the WISE \citep{wem+10}, as shown in
Section 2. Details for identifying galaxy clusters and evaluating
their physical parameters are given in Section 3. The comparisons with
previously known clusters in the optical, X-ray, and SZ cluster
catalogs are presented in Section 4. Cluster properties such as
dynamical states and the stellar mass of the brightest cluster
galaxies (BCGs) are discussed in Section 5. A summary is given in
Section 6.

Throughout this paper, we assume a flat Lambda cold dark matter
cosmology taking $H_0=70$ km~s$^{-1}$ Mpc$^{-1}$, $\Omega_m=0.3$, and
$\Omega_{\Lambda}=0.7$. All magnitudes are given in the AB system.

\begin{figure*}
\centering \includegraphics[width = 0.95\textwidth]{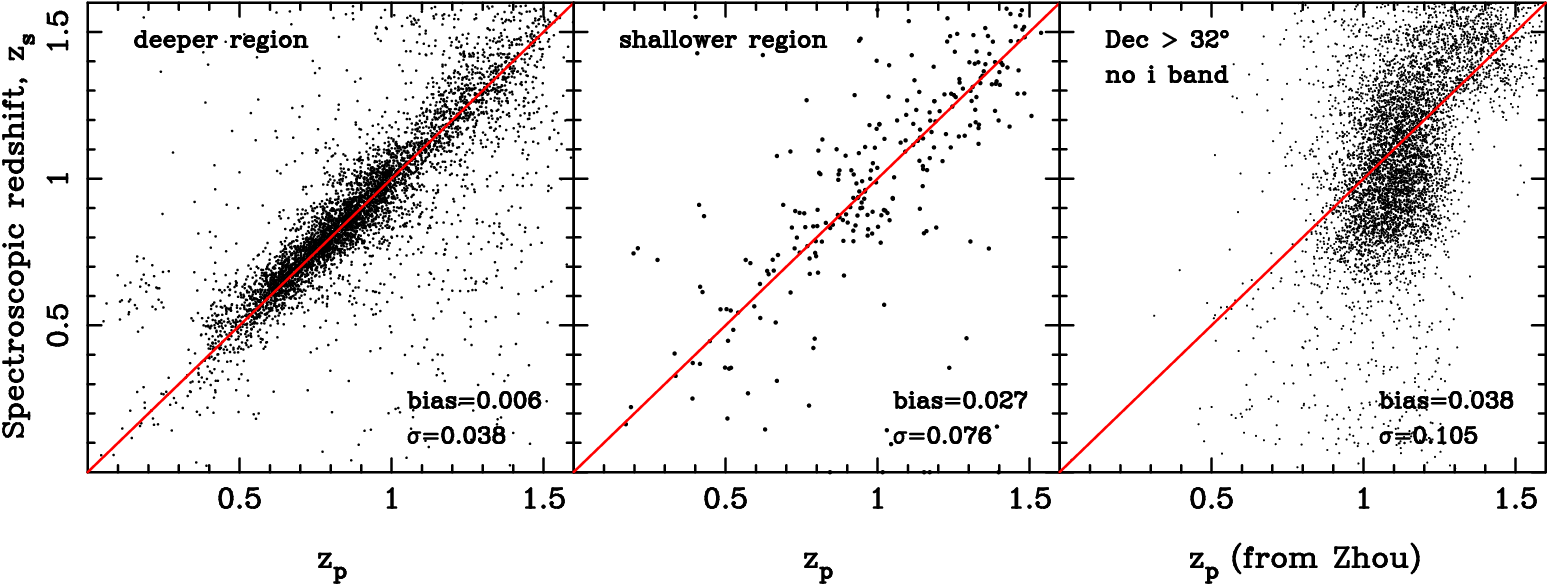}  
\caption{Comparison of photometric redshifts $z_{\rm p}$ with
    spectroscopic redshifts $z_{\rm s}$ for the faint galaxies of
    $z_{\rm m}>22$ in the deeper region ({\it the left panel}) and
    galaxies in the shallower region ({\it the middle panel}). {\it
      The right panel} is plotted for the galaxies in the shallower
    region of ${\rm Dec.}\ge32^{\circ}$ without $i$-band data, and
    the values of $z_{\rm p}$ are directly taken from
    \citet{znm+21}. The uncertainty and systematic bias are marked in
    each subpanel.}
\label{photozmag22}
\end{figure*}

\section{Survey data for galaxies}

The DESI Legacy Imaging Surveys \citep{dsl+19} consist of three
projects including the Dark Energy Camera Legacy Survey (DECaLs), the
Beijing-Arizona Sky Survey (BASS), and the Mayall $z$-band Legacy
Survey (MzLS).  The BASS and MzLS observe the northern hemisphere of
${\rm Dec.}\ge32^{\circ}$ with the survey magnitude depths of
$g=24.0$, $r=23.5$ and $z_{\rm m}=22.9$, respectively, for point
sources. The DECaLs covers the south hemisphere of ${\rm
  Dec.}<32^{\circ}$ with the magnitude depths of $g=24.5$, $r=23.9$
and $z_{\rm m}=22.9$, respectively, deeper than the northern part
\citep{zgz+19}. The southern survey part also includes the data from
the Dark Energy Survey \citep[DES][]{desdr2} with magnitude depths of
$g=25.3$, $r=25.0$ and $z_{\rm m}=23.9$, respectively, the data from
DECam Local Volume Exploration survey \citep{dfa+22} with the depth of
$z_{\rm m}=22.8$ and the data from DECam eROSITA
survey\footnote{https://noirlab.edu/science/programs/ctio/instruments/Dark-Energy-Camera/DeROSITAS}
with the depth of $z_{\rm m}=23.2$. The survey data in the DES region
are deepest, have a middle depth in the north Galactic cap region with
$-9.5^{\circ}\lesssim {\rm Dec.}<32^{\circ}$, and are shallowest in
the rest sky regions. The DESI Legacy Surveys data are supplemented
with the mid-infrared co-added imaging data from the unWISE
\citep{lang14,smg+19} with magnitude depths of ${\rm W1}=20.4$ and
${\rm W2}=19.5$, respectively.

The DESI Legacy Surveys DR9 contains the photometric data in the $grz$ and
${\rm W1W2}$ bands covering a sky area of about 19,700 deg$^2$. The
DR10 published data for objects from the DECaLs in the newly observed
regions and also new $i$-band data for objects in some southern parts
of DR9, covering a sky area of about 15,300 deg$^2$ with six bands
($griz{\rm W1W2}$). Current DESI Legacy Surveys have released data in
a total sky area of about 24,000 deg$^2$.

We analyze the DR9 and DR10 data independently for cluster
identification in this work because of the different number of bands
and choose the galaxy-type sources with ${\rm TYPE}\neq {\rm PSF}$.
The sources with bad photometry are discarded by using the
following quality flags \citep{yxh+21}:
${\rm FRAFLUX} < 0.5$, ${\rm FRACMASKED} < 0.4$ and ${\rm FRACIN} > 0.3 $
in both the $grz$ for the DR9 data and the $griz$ for the DR10
data. We also set a detection signal-to-noise of ${\rm S/N}\ge5$ in
the $grz{\rm W1}$ for the DR9 galaxies and $griz{\rm W1}$ for the DR10
galaxies. Some classified galaxies by their morphology are faint
contaminated stars, which can be partly removed in the color-color
space \citep{zdn+23} by setting
 $ r-z_{\rm m} \ge 1 $ and 
 $ z_{\rm m}-{\rm W1} > 1.2\,(r-z_{\rm m})-1.5. $
Even so, some faint stars of $r-z_{\rm m}<1$ can not be distinguished
from low redshifts galaxies. Fortunately they will not affect our
cluster identification work because we use only massive galaxies and
there are only a small number of such faint contaminated objects.

\subsection{Spectroscopic redshifts of galaxies}

Redshift of galaxies is a fundamental parameter for cluster
identification in our algorithm \citep{wh21,wh22}. We adopt
spectroscopic redshifts of galaxies if they are available in the Two
Micron All-Sky Survey (2MASS) Redshift Survey \citep{2mass}, the SDSS
DR17 \citep{sdssdr17} and the early data release of the DESI
\citep{desi23}. These three redshift surveys give spectroscopic
redshifts for 3.1 million galaxies in the DESI Legacy Surveys, which
provide accurate redshifts for a large number of member galaxies for
cluster identification.

\begin{deluxetable*}{rrccrcccccccccccccc}[t]
\tabletypesize{\scriptsize}
\tablecolumns{19}
\tablewidth{0pc}
\setlength{\tabcolsep}{0.9mm}
\tablecaption{The magnitudes, the estimated photometric redshifts
  and stellar masses for galaxies in the DESI Legacy Surveys DR10.
  The whole catalog is available at http://zmtt.bao.ac.cn/galaxy\_clusters/.}
\tablehead{ 
\colhead{R.A.} & \colhead{Dec.} &
\colhead{$g$} &  \colhead{${\rm Err}_g$} & \colhead{$r$} &  \colhead{${\rm Err}_r$} & \colhead{$i$} &  \colhead{${\rm Err}_i$} &
\colhead{$z_{\rm m}$} &  \colhead{${\rm Err}_z$} & \colhead{${\rm W1}$} &  \colhead{${\rm Err}_{\rm W1}$} & \colhead{${\rm W2}$} &  \colhead{${\rm Err}_{\rm W2}$} &
\colhead{$z_{\rm p}$} & \colhead{$\sigma_{\rm zp}$} & \colhead{$z_{\rm s}$}  & \colhead{$\log (M_\star)$} & \colhead{Flag}\\
  \colhead{(1)} & \colhead{(2)} & \colhead{(3)} & \colhead{(4)} & \colhead{(5)} & \colhead{(6)} & \colhead{(7)} & \colhead{(8)} &
  \colhead{(9)} & \colhead{(10)} & \colhead{(11)} & \colhead{(12)} & \colhead{(13)} & \colhead{(14)} & \colhead{(15)} &
  \colhead{(16)} & \colhead{(17)} & \colhead{(18)} & \colhead{(19)} 
}
\startdata 
0.00002 &  $-3.25165$ & 23.427 & 0.082& 22.989 & 0.069& 22.733 & 0.104& 22.274 & 0.105& 21.287 & 0.182& 22.241 & 0.755 & 1.2361 & 0.0774& $-1.0000$ & 10.01&0\\
0.00002 &  $-0.27648$ & 22.125 & 0.022& 20.862 & 0.008& 20.376 & 0.010& 19.980 & 0.011& 18.289 & 0.050& 18.404 & 0.050 & 0.4243 & 0.0976& $-1.0000$ & 10.74&0\\
0.00002 &  $-4.90665$ & 23.219 & 0.049& 22.688 & 0.039& 22.158 & 0.041& 21.682 & 0.040& 20.421 & 0.081& 20.590 & 0.205 & 1.1038 & 0.0791& $-1.0000$ & 10.33&0\\
0.00004 &  $-1.09510$ & 25.457 & 0.324& 23.719 & 0.082& 22.701 & 0.074& 21.971 & 0.052& 20.371 & 0.081& 20.360 & 0.172 & 0.9334 & 0.0540& $-1.0000$ & 10.44&0\\
0.00004 &  $-0.29959$ & 22.046 & 0.021& 21.152 & 0.012& 20.969 & 0.019& 20.709 & 0.024& 19.690 & 0.051& 20.880 & 0.284 & 0.4200 & 0.0695& $-1.0000$ &  9.87&0\\
0.00004 &  $-0.62148$ & 27.064 & 1.062& 24.269 & 0.152& 23.076 & 0.098& 22.659 & 0.107& 20.965 & 0.133& 21.070 & 0.310 & 0.7433 & 0.0578& $-1.0000$ & 10.02&0\\
0.00007 &  $-1.58000$ & 24.180 & 0.100& 23.386 & 0.058& 22.645 & 0.067& 22.463 & 0.078& 21.247 & 0.170& 23.265 & 1.362 & 0.7738 & 0.0250& $-1.0000$ &  9.55&0\\
0.00007 &  $-4.43194$ & 24.089 & 0.127& 23.726 & 0.112& 23.331 & 0.129& 22.532 & 0.102& 20.560 & 0.094& 20.393 & 0.180 & 1.4287 & 0.0960& $-1.0000$ & 10.65&0\\
0.00007 &  $-4.18086$ & 23.170 & 0.062& 22.282 & 0.033& 21.400 & 0.023& 21.063 & 0.028& 19.550 & 0.050& 20.062 & 0.136 & 0.8275 & 0.0357& $-1.0000$ & 10.55&0\\
0.00008 &  $-2.42179$ & 24.655 & 0.142& 23.533 & 0.065& 22.633 & 0.067& 22.079 & 0.057& 20.779 & 0.119& 20.912 & 0.285 & 0.8729 & 0.0767& $-1.0000$ & 10.15&0\\
\enddata
\tablecomments{
Descriptions of columns.
Columns 1 and 2: Right Ascension (R.A. J2000) and Declination (Dec. J2000) of a galaxy (in degree);
Columns 3--14: magnitudes and errors (AB system) in the $griz{\rm W1}{\rm W2}$ bands; 
Columns 15 and 16: photometric redshift and error;
Column 17: spectroscopic redshift, $-1.0000$ means not available;
Column 18: logarithm of stellar mass with $M_\star$ in unit of $M_\odot$; 
Column 19: flag for BCG-like galaxy, ``1" for yes, ``0" for no.
}
\label{tab1}
\end{deluxetable*}

\subsection{Photometric redshifts of galaxies}

The DESI Legacy Surveys DR9 includes photometric redshifts of galaxies
estimated by the random-forest algorithm using the $grz{\rm W1}{\rm
  W2}$ magnitude data \citep{znm+21}. We adopt these photometric
redshifts for 361.1 million galaxies in the DESI Legacy
Surveys DR9.

Following our previous papers \citep{wh21,wh22}, we estimate
photometric redshifts for galaxies in DR10 by using the
nearest-neighbor algorithm, which is based on the empirical relation
between galaxy colors and spectroscopic redshifts from a training
sample. In the color spaces, the close neighbors of galaxies have a
similar redshift. We calculate the distances in the color spaces
between a targeted galaxy and all galaxies in the training sample.
The photometric redshift of a targeted galaxy, $z_{\rm p}$, is
estimated to be the median spectroscopic redshift of the 20 nearest
neighbors. The uncertainty of the photometric redshift, $\sigma_{\rm
  z_p}$, is taken as the dispersion of these 20 spectroscopic
redshifts.

\begin{figure}
\centering \includegraphics[width = 0.8\columnwidth]{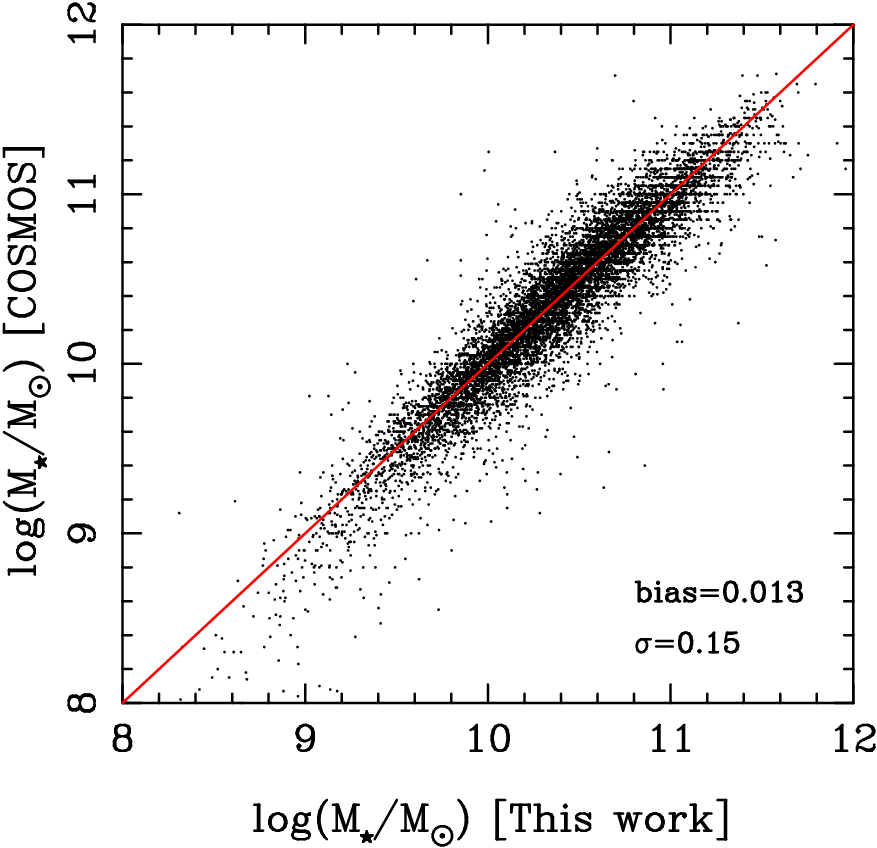}
\caption{Comparison between our estimated stellar masses of galaxies
  with the values in the COSMOS2015 catalog.}
\label{stellarmass}
\end{figure}

\begin{figure}
\centering \includegraphics[width = 0.9\columnwidth]{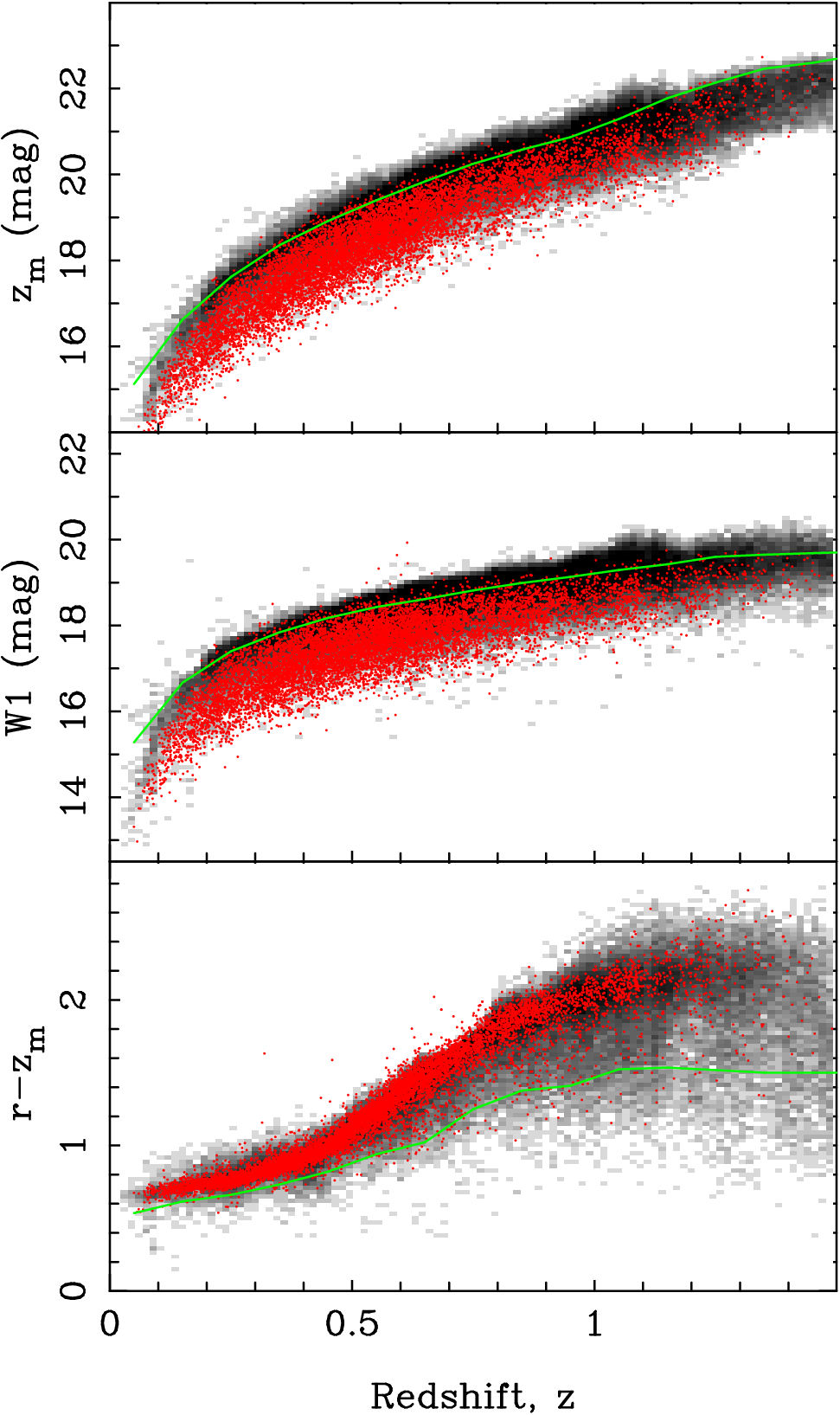}
\caption{Magnitudes at $z$ and ${\rm W1}$ bands and color ($r-z_{\rm
    m}$, bottom) against redshifts for 10,000 randomly-selected known
  BCGs, as indicated by small (red) dots. The distribution of all
  massive galaxies of $M_\star\ge 10^{11}~M_{\odot}$ in the DESI
  Legacy Surveys are shown with the number density in gray.  We
  set criteria for BCG-like galaxies as shown by the solid line.}
\label{BCGmag}
\end{figure}

For this work, we get a training sample of 883,000 galaxies with
spectroscopic redshifts from the compiled data in \citet{wh21} and
\citet{wh22}, and also galaxies from the flux-limited data in the
Galaxy And Mass Assembly Survey \citep[GAMA,][]{lbd+15,dbr+22}, 
VIPERS DR2 \citep{sgg+18} and DEEP2 \citep{mnc+13}, plus the local
galaxies in the 2MASS Redshift Survey \citep{2mass} and the galaxies
in the early data release of the DESI \citep{desi23}. The galaxies in
this sample has a redshift in the range of $z<1.6$.

We then estimate photometric redshifts for 295.6 million galaxies in
the DESI Legacy Surveys DR10 with $griz{\rm W1}{\rm W2}$ magnitude
data, see Table~\ref{tab1}. In Figure~\ref{photoztest}, we compare the
photometric redshifts with spectroscopic redshifts for galaxies with
different $z$-band magnitudes in the region of middle survey
depth. For bright galaxies of $z_{\rm m}<21$, the uncertainty of
photometric redshifts, defined as $\sigma_{\Delta z}=1.48\times {\rm
  median} (|z_p-z_s| /(1+z_s))$, is only about 0.013 for the galaxies
at $z<1$. The uncertainty increases to 0.028 for the galaxies of
$21<z_{\rm m}<21.5$ at $0.7<z<1.3$, and to 0.038 for the galaxies of
$21.5<z_{\rm m}<22$ at $0.9<z<1.5$. The photometric redshifts are
reasonably estimated even for very faint galaxies of $z_{\rm m}>22$,
with an uncertainty of only 0.053 and a fairly small systematic bias
of $-0.041$. The systematic bias and uncertainty are larger for
fainter galaxies at $z>1.2$, which are probably caused by the small
number of faint galaxies at high redshifts in the training
sample. Moreover, the 4000\,\AA break is moving out of the $z_{\rm m}$
band for objects of $z>1.2$, making the estimate of photometric
redshifts difficult. The survey depth does affect the accuracy of
photometric redshifts, especially for faint galaxies. As seen in
Figure~\ref{photozmag22}, the photometric redshift is slightly more
accurate in the deeper region and worse in the shallower region for
galaxies of $z_{\rm m}>22$.

Based on the DESI Legacy Surveys DR9 data, \citet{znm+21} used the
$grz{\rm W1}{\rm W2}$ magnitudes to estimate photometric
redshifts. For galaxies of $z_{\rm m}<21$, the redshifts are well
estimated with an uncertainty of about 0.01332, but not for fainter
galaxies. For the galaxies of $z_{\rm m}>22$ in the shallow region of
${\rm Dec.}\ge32^{\circ}$, the photometric redshifts are
systematically overestimated at $z_{\rm s}<1$ and underestimated at
$z_{\rm s}>1$, with an uncertainty as large as 0.105 (see the right
panel of Figure~\ref{photozmag22}). We have tested our algorithm by
ignoring the newly added $i$-band magnitude data for galaxies in the
DR9 and DR10 and get the consistent results with \citet{znm+21}. We
therefore conclude that the newly added $i$-band data in the DR10 can
significantly improve the accuracy of photometric redshift estimates.

In summary, we take photometric redshifts for 361.1 million galaxies
from the DESI Legacy Surveys DR9, and also the newly estimated
photometric redshifts for 295.6 million galaxies from the DR10 for
cluster identification.

\subsection{Stellar masses of galaxies}

Stellar masses of galaxies are another fundamental parameter for
cluster identification in our algorithm. The optical and infrared
luminosities of galaxies are good tracers of their stellar mass
\citep{bmk+03,khw+03,wwz+13}. In \citet{wh21}, we took the values from
the COSMOS2015 catalog \citep{lmi+16} to calibrate the relation
between the ${\rm W1}$-band luminosity, the color $r-z_{\rm m}$, and
the stellar mass of galaxies. Here, we do a similar calibration but
for the relation of the stellar mass with the $z$-band luminosity. We
then obtain two kinds of stellar mass estimates for galaxies from the
${\rm W1}$ and $z$-band luminosities separately and then take the
average. The final values are listed in Column (18) of
Table~\ref{tab1}. Compared to the stellar masses of galaxies in the
COSMOS2015 catalog, our newly estimated masses have an uncertainty of
only 0.15 dex (Figure~\ref{stellarmass}), much improved from the
single ${\rm W1}$ band \citep{wh21}.

\begin{deluxetable*}{crrrcccccrccrcccl}
\rotate
\tablecaption{A catalog of 1.58 million clusters of galaxies identified from the
  DESI Legacy Surveys.}
\tablewidth{1pt}
\tablehead{ 
  \colhead{ID} & \colhead{Name} & \colhead{R.A.} & \colhead{Dec.} & \colhead{$z_{\rm cl}$} & \colhead{flag$_z$} &
  \colhead{$z_{\rm m,BCG}$} & \colhead{${\rm W1_{BCG}}$} &  \colhead{$\log(M_{\star,\rm BCG})$} & \colhead{$r_{500}$} &
  \colhead{$\lambda_{500}$} & \colhead{$M_{500}$} & \colhead{$N_{\rm gal}$} & \colhead{$\Gamma$} & \colhead{${\rm \sigma}_{\Gamma}$}
  & \colhead{Source} & \colhead{Other catalog} \\
\colhead{(1)} & \colhead{(2)} & \colhead{(3)} & \colhead{(4)} & \colhead{(5)} & 
\colhead{(6)} & \colhead{(7)} & \colhead{(8)} & \colhead{(9)} & \colhead{(10)} &
\colhead{(11)} & \colhead{(12)} & \colhead{(13)} & \colhead{(14)} & \colhead{(15)} & \colhead{(16)} & \colhead{(17)}
}
\startdata 
 1&    J000000.1$-$372835 & 0.00046 &$-37.47628$ & 0.1364 & 0& 15.377& 15.561 & 11.49& 0.766 & 25.79 & 1.17 & 14 &  9.99 &  9.99 & 1& WHY18    \\
 2& WH-J000000.3$-$563518 & 0.00135 &$-56.58835$ & 0.1957 & 0& 15.712& 15.723 & 11.70& 0.625 & 23.24 & 1.06 &  7 &  9.99 &  9.99 & 1&          \\
 3&    J000000.4$-$391925 & 0.00172 &$-39.32369$ & 0.3580 & 0& 17.713& 17.313 & 11.41& 0.526 & 15.90 & 0.73 &  7 &  9.99 &  9.99 & 1& Y21      \\
 4&    J000000.5$+$021911 & 0.00201 &$  2.31980$ & 0.4282 & 1& 18.043& 17.359 & 11.55& 0.612 & 23.40 & 1.07 & 13 &  9.99 &  9.99 & 1& redMaPPer\\
 5& WH-J000000.6$-$590047 & 0.00260 &$-59.01297$ & 0.1380 & 0& 14.993& 15.186 & 11.69& 0.606 & 15.71 & 0.73 &  7 &  9.99 &  9.99 & 1&          \\
 6&    J000000.8$-$412726 & 0.00327 &$-41.45719$ & 0.4015 & 0& 18.386& 17.691 & 11.29& 0.462 & 10.95 & 0.51 &  6 &  9.99 &  9.99 & 1& CFSFDP   \\
 7&    J000001.0$+$012416 & 0.00403 &$  1.40456$ & 0.6538 & 1& 18.611& 17.088 & 11.85& 0.560 & 24.77 & 1.13 &  6 &  9.99 &  9.99 & 1& CFSFDP   \\
 8& WH-J000001.0$-$594933 & 0.00426 &$-59.82570$ & 0.6250 & 0& 19.248& 18.155 & 11.29& 0.457 & 15.18 & 0.70 &  8 &  9.99 &  9.99 & 1&          \\
 9&    J000001.0$-$535824 & 0.00435 &$-53.97320$ & 0.6112 & 0& 18.693& 17.541 & 11.66& 0.607 & 23.42 & 1.07 & 11 &  9.99 &  9.99 & 1& redMaPPer\\
10&    J000001.2$-$030053 & 0.00504 &$ -3.01485$ & 0.5090 & 1& 18.241& 17.339 & 11.67& 0.543 & 21.71 & 0.99 &  8 &  9.99 &  9.99 & 1& Y21      \\ 
\enddata
\tablecomments{
Column 1: Cluster ID;
Column 2: Cluster name with J2000 coordinates. Names with `WH' are newly identified in this paper;    
Columns 3 and 4: Right Ascension (R.A. J2000) and Declination (Dec. J2000) of cluster BCG (in degree);
Column 5: cluster redshift $z_{\rm cl}$;
Column 6: redshift flag, '0' for photometric redshift in Column 5 and '1' for spectroscopic redshift;
Columns 7--8: BCG magnitudes (AB system) in the $z$ and ${\rm W1}$ bands, respectively;
Column 9: logarithm of BCG stellar mass with $M_{\star,\rm BCG}$ in unit of $M_{\odot}$;
Column 10: cluster radius, $r_{500}$, in Mpc; 
Column 11: cluster richness;
Column 12: derived cluster mass, in units of $10^{14}~M_{\odot}$;
Column 13: number of member galaxy candidates within $r_{500}$;
Columns 14--15: relaxation parameter and error for rich clusters with $N_{\rm gal}\ge30$, otherwise 9.99 is given;
Column 16: data source for the cluster. '1' for the data with $i$-band magnitude available and '2' for the data without $i$-band magnitude;
Column 17: Reference notes for previously known clusters: maxBCG \citep{kma+07}, WHL \citep{whl09,whl12,wh15}, GMBCG \citep{hmk+10},
AMF \citep{spd+11,bsp+18}, redMaPPer \citep{rrb+14,rrh+16}, CAMIRA \citep{ogu14}, WHY18 \citep{why18},  
WH21 \citep{wh21}, Y21 \citep{yxh+21}, WaZP \citep{abd+21}, WH22 \citep{wh22}, CFSFDP \citep{zgx+21,zsx+22}, CluMPR \citep{ynd+23}.\\
(This table is available in its entirety in a machine-readable form.)
}
\label{tab2}
\end{deluxetable*}

\section{Galaxy clusters identified from the DESI Legacy Surveys}

In optical, galaxy clusters show a distinct overdensity of galaxies
around a massive central galaxy. The total stellar mass of member
galaxies is tightly related to cluster mass
\citep{and10,pnh+18,pab+20}. Using the photometric redshifts, we can
select most of the member galaxies with a small fraction of
contamination \citep{whl09}. Therefore, we identify galaxy clusters by
searching for the overdensity of stellar mass within a photometric
redshift slice around massive galaxies.

\subsection{Pre-selection for BCG-like massive galaxies}

Galaxies clusters generally contain one or more massive galaxies, and
the central one is called the BCG. Pre-selection for BCG-like galaxies
can help to identify galaxy clusters more efficiently.

As the most massive galaxies in the Universe, the BCGs are very
luminous in the optical and infrared bands \citep{hsw+09}. In general,
they contain an old stellar population and show a red color
\citep{wad+08}, and are located at the bright end in the
magnitude--redshift diagram and at the red end in the color--redshift
diagram \citep{why18}. We select BCG-like massive galaxies based on
the $rz{\rm W1}$ magnitudes and the estimated redshift and stellar
mass.

First, the BCG-like galaxies are selected to have a stellar mass of
$M_\star\ge 10^{11}~M_{\odot}$ \citep{lsm+12,wh21}. Then, according to
the distributions of known BCGs from Abell clusters \citep{lps+14},
WHL and WH22 clusters \citep{whl12,wh15,wh22}, we set criteria shown
in Figure~\ref{BCGmag}, which cause a lost of 2.5\% of the known faint
BCGs but exclude many fainter galaxies detected in the $z$ and ${\rm
  W1}$ bands. Similarly, some blue massive galaxies are also discarded
according to the threshold shown in the color-redshift diagram
($r-z_{\rm m}$) in Figure~\ref{BCGmag}. Note that BCGs at low
redshifts has a tight distribution in colors, but these at high
redshifts have much scattered colors due to star formation
\citep{obp+08, lmm12, msb+16,wh21}.

In total, 15.23 million galaxies are pre-selected as possible BCGs, as
indicated in Column (19) of Table~\ref{tab1}, of which 9.40 million
galaxies have $i$-band data.

\subsection{Identification of galaxy clusters}

Following our previous works \citep{wh21,wh22}, we identify galaxy
clusters by searching for the overdensity of stellar mass around massive
galaxies within a redshift slice and then estimating the
parameters of cluster candidates.

Previously, to get cluster radius and mass for to-be-identified galaxy
clusters, we first calibrated the scaling relation between the total
stellar mass within a given radius and cluster radius ($r_{500}$) or mass ($M_{500}$)
by using a sample of $M_{500}$ known clusters. Here, $r_{500}$ is the
radius within which the mean density is 500 times the critical density
of the universe. $M_{500}$ is the cluster mass within $r_{500}$. We
took galaxies within a photometric redshift slice, $z\pm\Delta z$,
around massive galaxies as member galaxy candidates for cluster
candidates. The half of the slice thickness is taken as being
\begin{equation}
  \Delta z=\left\{
\begin{array}{ll}
   0.04\,(1+z)           &     \mbox{for $z\leq 0.7$}\\
   0.15\,z-0.037       &     \mbox{for $z> 0.7$}
\end{array}
\right. \,.
\label{pzslice}
\end{equation}
We discard the member galaxy candidates with a large redshift
uncertainty of ${\sigma_{zp}} > 2\,\Delta z$. Then, we searched for
galaxy cluster candidates from the whole photometric data. The
redshift of the cluster candidate was estimated as the median value of
photometric redshifts of member galaxy candidates. The cluster radius
and richness (a mass proxy) were estimated according to the total
stellar mass of member galaxy candidates and the calibrated scaling
relations. A galaxy cluster is identified if it is above a richness
threshold.

\begin{figure}
\centering \includegraphics[width = 0.95\columnwidth]{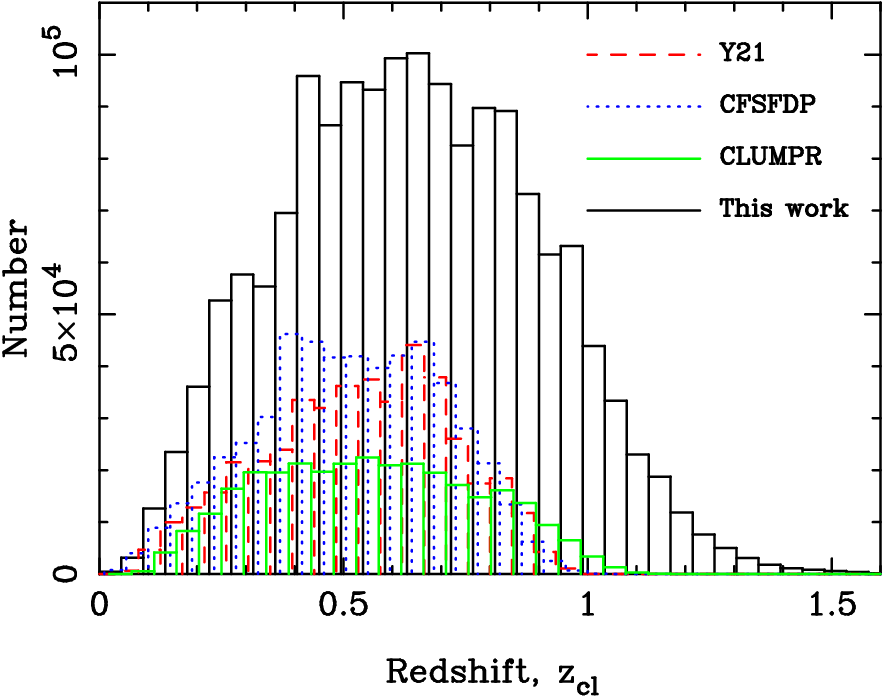}
\caption{Redshift distribution of the identified 1,581,179
  clusters, compared to clusters previously identified from DESI
  Legacy Surveys DR9 including Y21 \citep{yxh+21}, CFSFDP \citep{zsx+22}
  and CLUMPR \citep{ynd+23}).}
\label{z-dis}
\end{figure}

\begin{figure*}[t]
\centering \includegraphics[width = 0.75\textwidth]{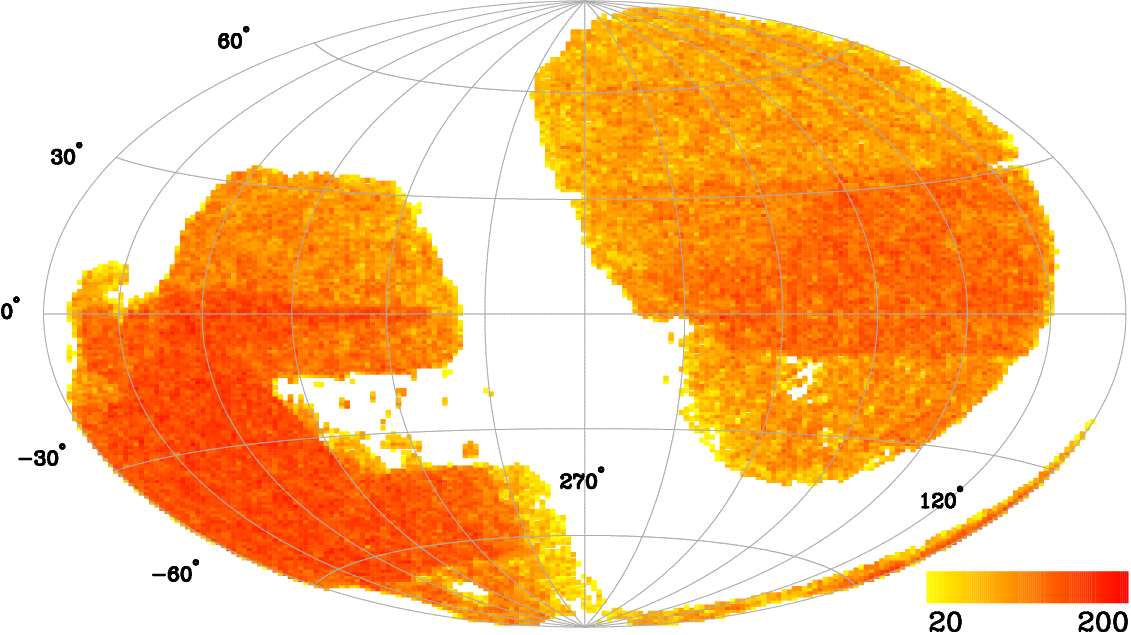}\\[2mm]
\caption{The sky distribution of identified galaxy clusters in the
  Equatorial coordinate system, expressed by the number of clusters
  per square degree.}
\label{skycov}
\end{figure*}

In this work, we make some improvements to our procedures. First, only
massive galaxies with $M_\star\ge 10^{10}~M_{\odot}$ are used since
they have a smaller redshift uncertainty than these less massive (and
fainter) galaxies. The central BCGs of to-be-identified clusters are
therefore only taken from the pre-selected BCG-like galaxies. Second,
the spectroscopic redshifts of some galaxies, if available, are used
to determine the redshifts of cluster candidates and to label member
galaxies. We adopt the available spectroscopic redshift of the BCG as
the spectroscopic redshift of a cluster. If the spectroscopic redshift
of the BCG is not available, the available spectroscopic redshifts of
other galaxies are adopted if they are within $0.025(1+z)$ from the
cluster photometric redshifts. The member galaxies with spectroscopic
redshifts are determined if they have a velocity difference $\Delta
v<2500$ km~s$^{-1}$ from the cluster spectroscopic redshifts. Thirdly,
we verify the scaling relations of $r_{500}$ and $M_{500}$ by using a
large cluster sample \citep{wh15}, in which the cluster masses have
been re-scaled for consistency and calibrated to weak lensing
measurements \citep{vbe+09}. As done in our previous \citep{wh21}, we
get the scaling relation for $r_{500}$ as being
\begin{eqnarray}
\log r_{500}&=&0.402\log \frac{M_{\star,r1}}{10^{10}M_\odot}-(0.944\pm0.03)\nonumber\\
&&+(0.43\pm0.02)\log (1+z),
\label{r500z}
\end{eqnarray}
where $M_{\star,r1}$ is the total stellar mass (after background
subtracted) of member galaxy candidates within a projected radius of
$r_1=1.0\,E(z)^{-1/3}~{\rm Mpc}$. The total stellar mass,
$M_{\star,500}$, is then calculated from galaxies within a projected
radius of $r_{500}$. We get the scaling relation between
$M_{\star,500}$ and cluster mass $M_{500}$, which is consistent with
our previous result \citep{wh21}. The richness of a galaxy cluster,
$\lambda_{500}$, is then defined as a redshift-independent
mass proxy by
\begin{equation}
  \lambda_{500}=M_{\star,500}(1+z)^{0.21}/M^{\ast}_\star,
\label{richdef}
\end{equation}
where $M^{\ast}_\star$ is a characteristic stellar mass. A galaxy
cluster is identified when it has a richness of $\lambda_{500}\ge10$ and
the number of member galaxy candidates within $r_{500}$ as being
$N_{\rm gal}\ge6$. The threshold is slightly lower than that we
adopted in \citet{wh21} so that we do find more low-mass clusters.

After cleaning the repeated entries of identified clusters, we finally
get 1,581,179 galaxy clusters, as listed in
Table~\ref{tab2}. Among them, 877,806 clusters are identified
for the first time; 946,486 clusters are identified by using our
newly obtained photometric redshifts, and the other 634,693
clusters are identified from photometric redshifts in the DESI Legacy
Surveys DR9 data.

The redshift distribution of clusters extends to $z\sim1.5$, and is
compared to those clusters previously found from the DESI Legacy
surveys DR9 \citep{yxh+21,zsx+22,ynd+23}, see Figure~\ref{z-dis}.

The sky distribution of galaxy clusters is shown in
Figure~\ref{skycov}. The clusters in the DES region in the south
Galactic cap has the highest number density as being $\sim$108
clusters per deg$^2$. The density declines to $\sim$80 clusters per
deg$^2$ in the north Galactic cap region with $-9.5^{\circ}\lesssim
{\rm Dec.}<32^{\circ}$, and to about $\sim$50 clusters per deg$^2$ in
the rest regions.

\begin{figure*}
  \centering
  \includegraphics[width = 0.8\columnwidth]{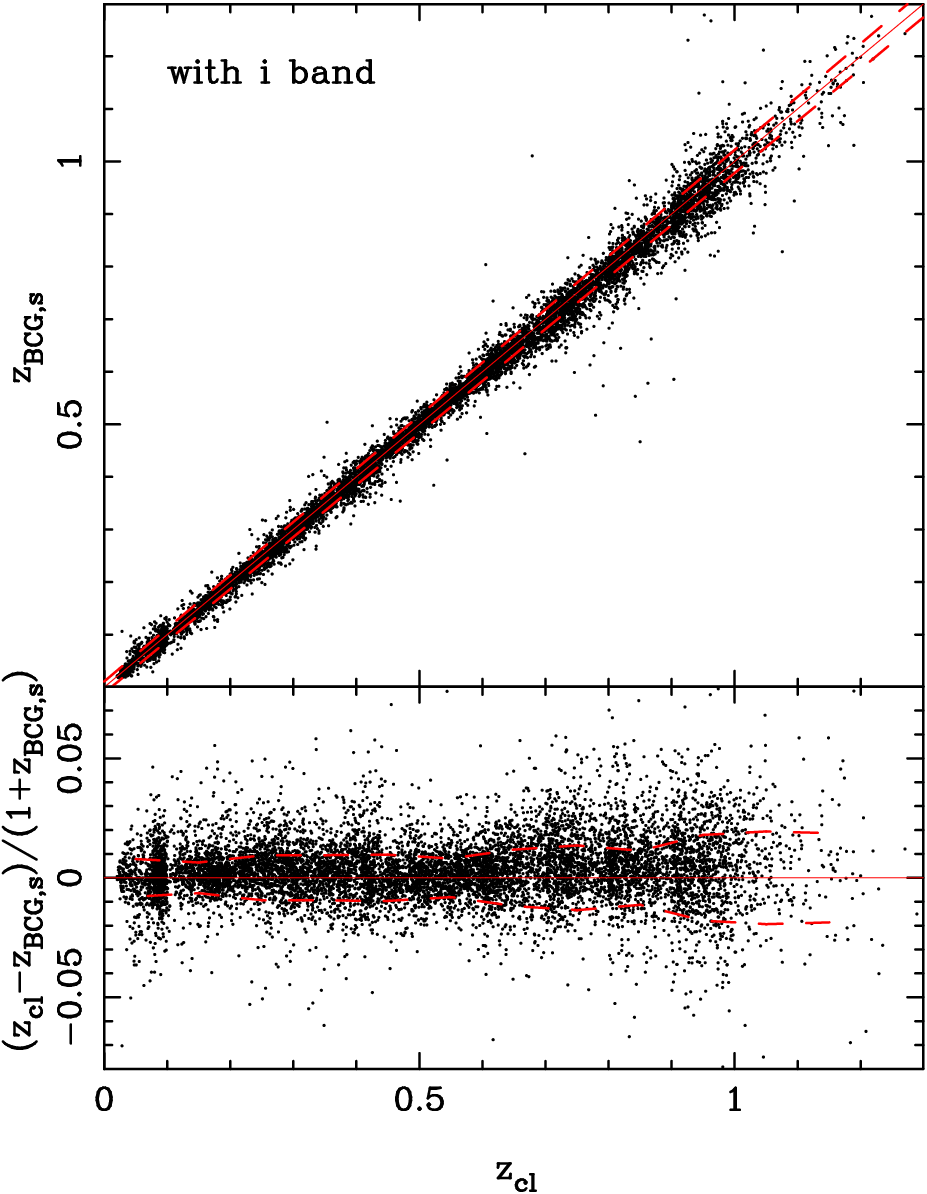}
  \includegraphics[width = 0.8\columnwidth]{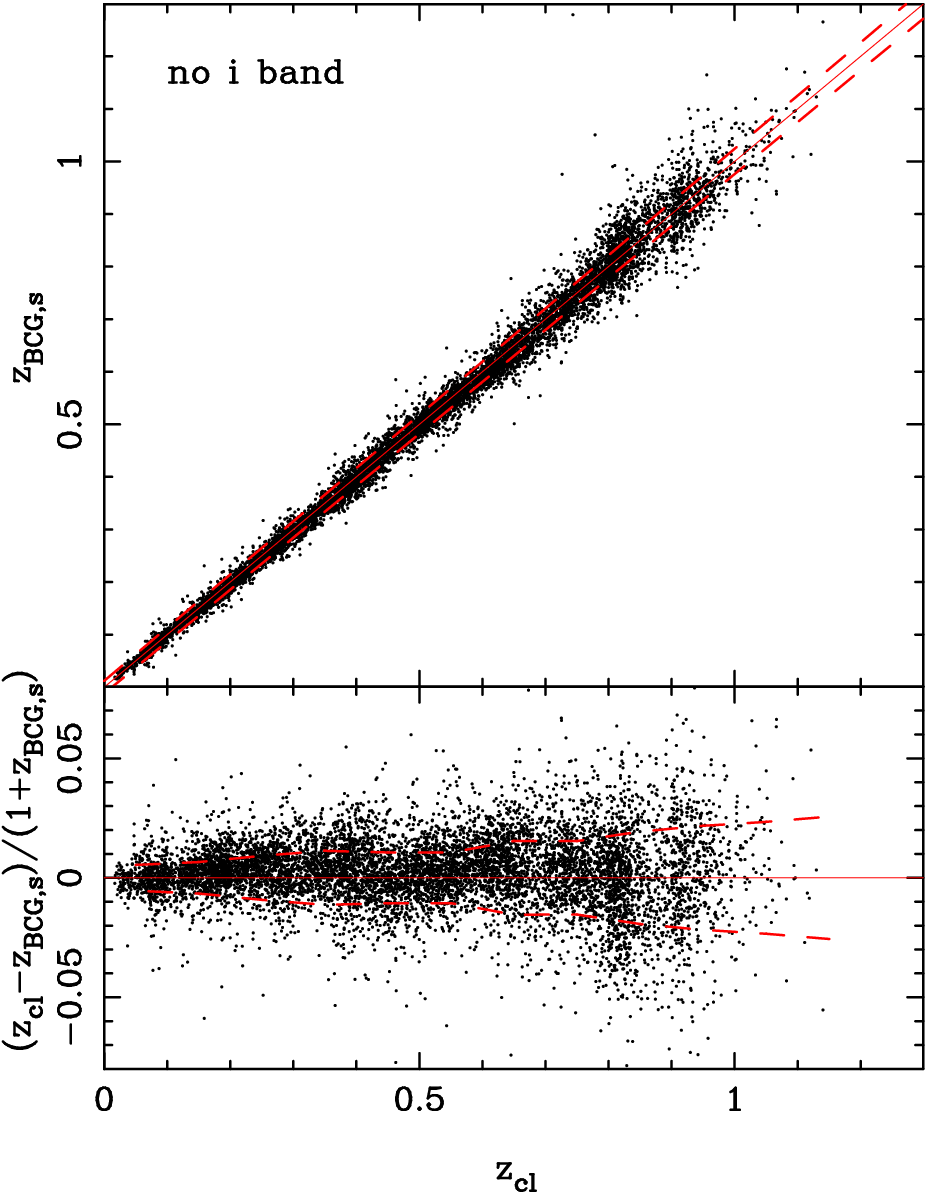}
\caption{In the left panel, the comparison between photometric
  redshifts and spectroscopic redshifts for clusters identified with
  $i$ band data. In every redshift interval of 0.1, 1000 clusters are
  randomly. The solid line is for equal values, and dashed lines
  represent the deviation of 1\,$\sigma$. The right panel is the same
  as the left, but for clusters identified without $i$ band data in
  the shallowest region of ${\rm Dec.}\ge32^{\circ}$.}
\label{pzc}
\end{figure*}

\begin{figure}
  \centering \includegraphics[width = 0.80\columnwidth]{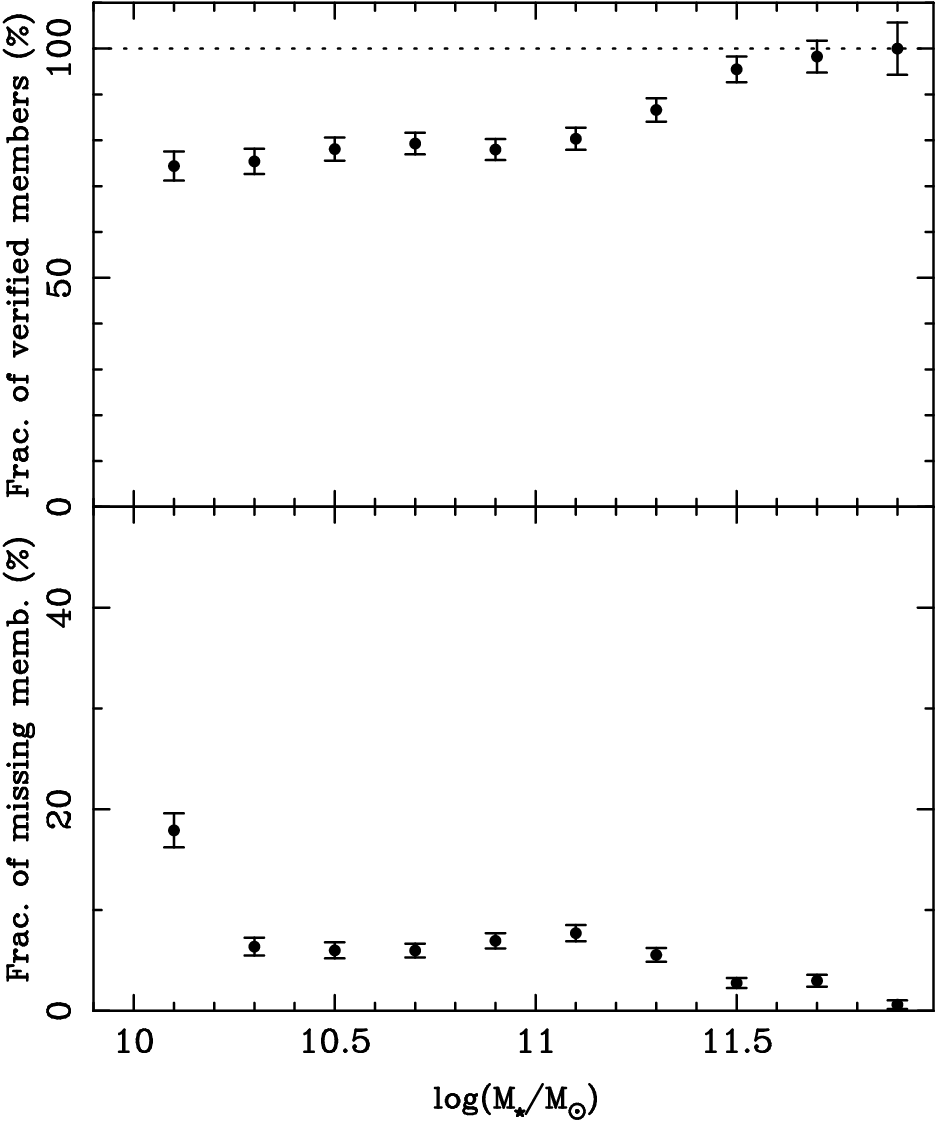}
\caption{Verification of cluster members as a function of the stellar
  mass of galaxies by using the spectroscopic survey data of the GAMA
  \citep{dbr+22}. Upper panel: the fraction of verified member
  galaxies ($N_{\rm veri.}/N_{\rm all}$); Lower panel: the fraction of
  missed true member galaxies $(N_{\rm missing}/ N_{\rm real})$. Here,
  $N_{\rm veri.}$, $N_{\rm missing}$, $N_{\rm real}$ are number of
  galaxies defined within a small spectroscopic redshift offset from
  the BCGs.}
\label{mem}
\end{figure}

\begin{figure}
\centering \includegraphics[width = 0.85\columnwidth]{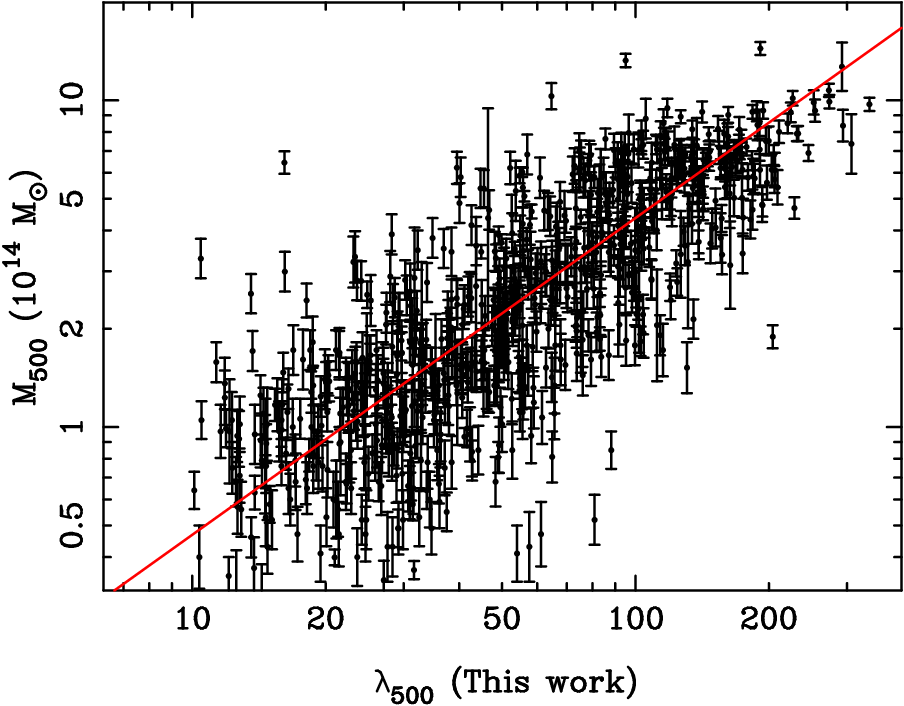}
\caption{Correlation between cluster richness and cluster masses compiled in
  \citet{wh15}. The solid is the best fit for the data.}
\label{massrich}
\end{figure}

\subsection{Verification of clusters and their parameters}
\label{verify}

To validate the reliability of identified clusters, we cross-match our
identified clusters in the deep COSMOS field with previously known
clusters. The COSMOS has a field of view of 2 deg$^2$ and more than
30-band deep imaging observations at the wavelengths from NUV
\citep[e..g.][]{zsr+07}, optical \citep[e.g.][]{saa+07,tsm+07} to
infrared \citep[e.g.][]{ssa+07}. Deep X-ray observations were also
carried out for this sky area by the Chandra and XMM-Newton
\citep[e.g.][]{hcb+07,cmc+16}. Based on these high-quality data,
galaxy clusters/groups have been identified with high completeness at
low redshifts \citep{kli+12,wh11,scc+12, wh15,dmm+17, gft+19}. Most of
the clusters in the field have spectroscopic observations.

In the COSMOS field, we detect 221 clusters, 192 of which have
counterparts in the previous catalogs. At $z<0.9$, only four of 145
clusters are new. We carefully inspect all 221 clusters in the DESI
color images and find that they all have a concentration of galaxies
around their BCGs with a similar color. We conclude that our cluster
sample has a high purity.

In our cluster sample, 338,841 clusters have spectroscopic redshifts
for one or more galaxies. To test the accuracy of cluster photometric
redshifts, we perform our algorithm by using photometric redshifts
only for these clusters. Figure~\ref{pzc} shows the comparison between
cluster photometric redshifts and spectroscopic redshifts. One can get
the uncertainty of photometric redshift, $\sigma^{\rm
  cl}_{z}=1.48\times {\rm median}(|z_{\rm cl}-z_{\rm s}|/(1+z_{\rm
  s}))=0.010$ for clusters identified from the data with available
$i$-band magnitudes (see the left panel of Figure~\ref{pzc}), which is
$\sigma^{\rm cl}_{z}=0.012$ for the clusters identified from the data
without $i$ band in the shallowest region of ${\rm Dec.}\ge32^{\circ}$
(see the right panel of Figure~\ref{pzc}).

Most member galaxy candidates are obtained by photometric redshifts,
except for those with spectroscopic redshifts. The accuracy and
completeness of the detected member galaxies are tested by using the
spectroscopic survey data of $\sim$300,000 galaxies from the GAMA
survey down to flux-limited of $r< 19.8$ mag covering $\sim$286
deg$^2$ \citep{dbr+22}. Those galaxies with a velocity difference less
than 2500 km~s$^{-1}$ from the BCGs are taken as true member
galaxies. The member galaxy candidates of clusters determined by
photometric redshifts only are cross-matched with the spectroscopic
data of the GAMA. We find that 78\% of the member galaxy candidates
can be verified by spectroscopic redshift if the BCGs are not
included, or up to 83\% if BCGs are included. As seen in
Figure~\ref{mem}, more massive galaxies have a higher fraction of true
members detected. Due to the uncertainty of photometric redshifts,
some true members are not included in the redshift slices.  The
missing member galaxies are such galaxies with a velocity difference
of less than 2500 km~s$^{-1}$ from the BCGs. The missing fraction is
about 8\% of true members, slightly higher for lower mass galaxies
(see Figure~\ref{mem}). Note that the GAMA survey is much shallow
compared to the DESI Legacy Surveys, this test is made only for the
bright galaxies in the DESI Legacy Surveys.

The cluster richness $\lambda_{500}$ is compared to the mass $M_{500}$
for the common clusters in \citet{wh15}. As shown in
Figure~\ref{massrich}, they are well correlated and can be scaled by
\begin{equation}
\log M_{500}=(0.97\pm0.03)\log \lambda_{500}-(1.30\pm0.02),
\label{m500rich}
\end{equation}
where $M_{500}$ is in unit of $10^{14}~M_{\odot}$. The richness
threshold of $\lambda_{500}=10$ corresponds to the mass of
$M_{500}\sim 0.47\times10^{14}~M_{\odot}$. By the
Equation~\ref{m500rich}, the $M_{500}$ can be estimated by
$\lambda_{500}$ with an uncertainty of 0.20 dex for all identified clusters.

\begin{figure*}
  \centering 
  \includegraphics[width = 0.33\textwidth]{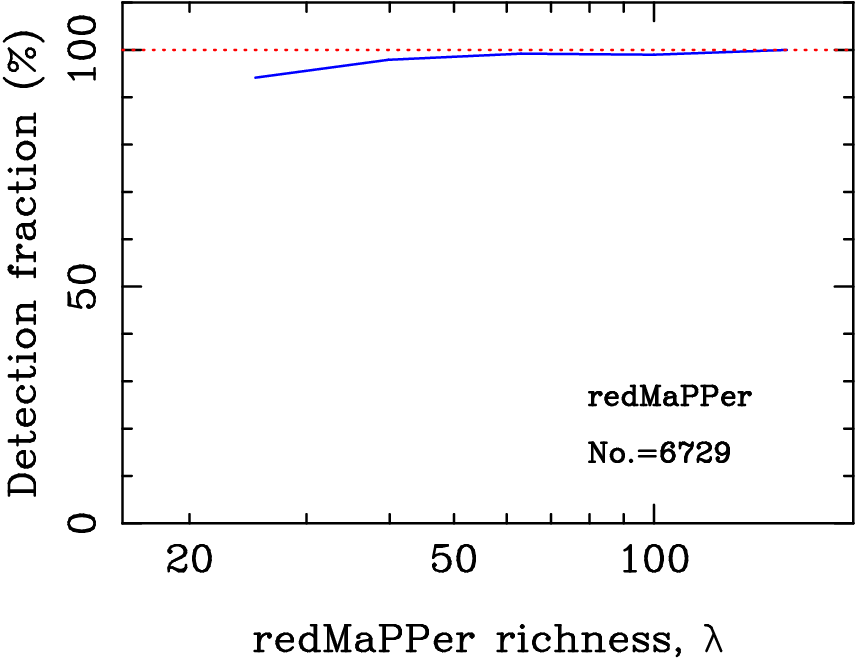} 
  \includegraphics[width = 0.33\textwidth]{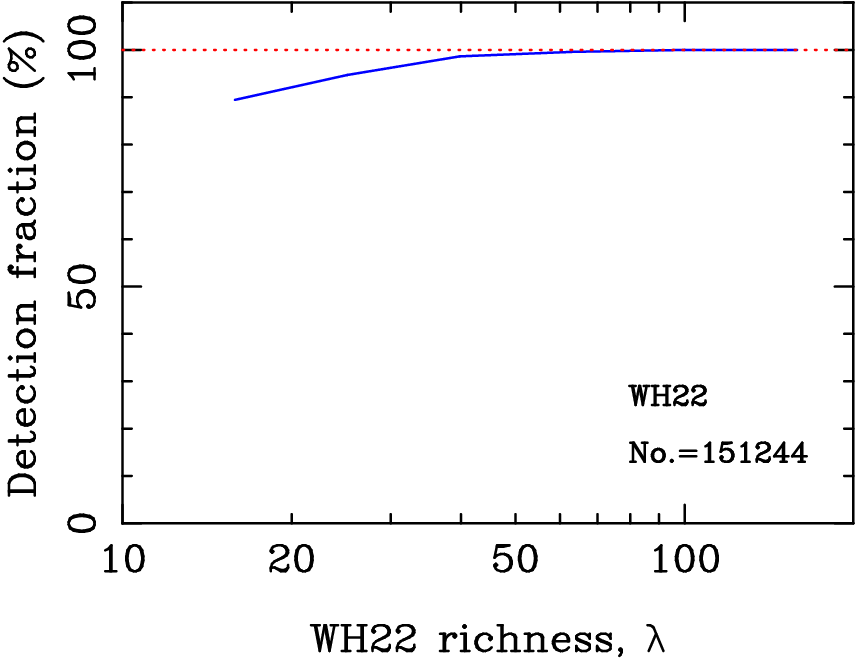} \\[3mm]
  \includegraphics[width = 0.32\textwidth]{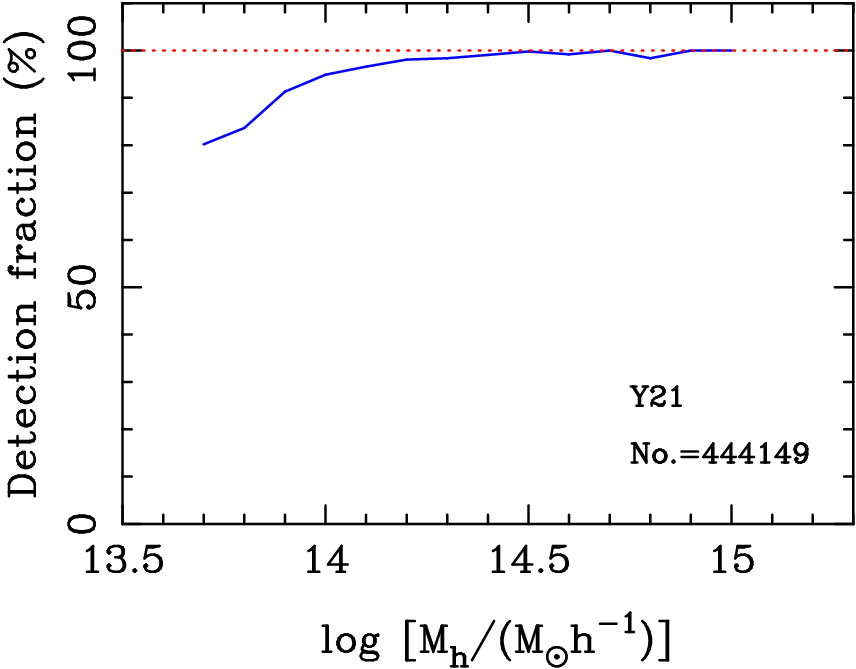}
  \includegraphics[width = 0.33\textwidth]{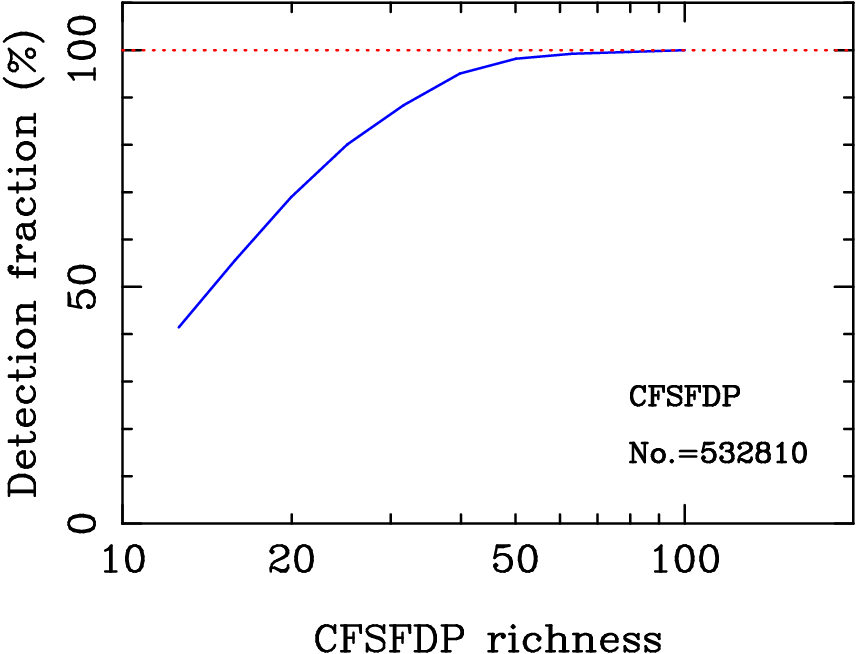} 
  \includegraphics[width = 0.33\textwidth]{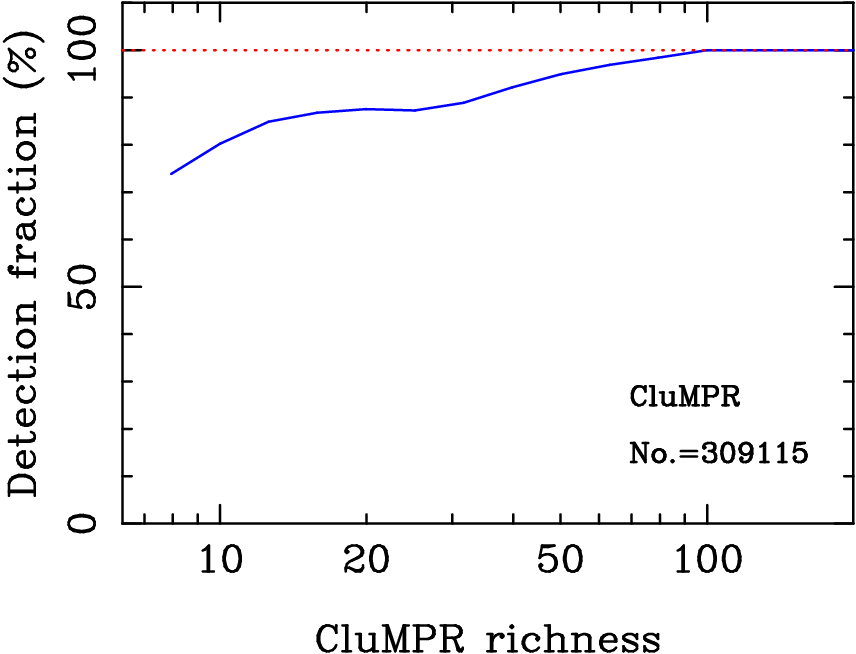}  
\caption{The detection fraction of clusters as a function of richness
  or halo mass for clusters in the redMaPPer \citep{rrh+16}, WH22
  \citep{wh22}, Y21 \citep{yxh+21}, CFSFDP \citep{zsx+22} and CluMPR
  \citep{ynd+23} catalogs.}
\label{comp}
\end{figure*}

\section{Cross-matching clusters with those in previous catalogs}

In the sky area of DESI Legacy Surveys, some survey data have
previously been released, from which a large number of clusters
  of galaxies have been identified in the literature, as marked in the
  last column of Table\ref{tab2}.

In the following, we cross-match the identified clusters with those in
some defined redshift ranges and sky regions in previous cluster catalogs,
including the optical clusters from the DESI Legacy surveys and DES,
the X-ray clusters from {\it ROSAT} All-Sky Survey (RASS) and extended
ROentgen Survey with an Imaging TelescopeArray (eROSITA), and the SZ
clusters from Planck, South Pole Telescope (SPT) and Atacama Cosmology
Telescope (ACT).

\subsection{DES and DESI Legacy Surveys clusters}

\citet{rrb+14} identified galaxy clusters based on the red-sequence
feature of cluster galaxies, and presented the redMaPPer cluster
catalog. The red sequence model was calibrated by using a training
sample and then applied to photometric data to search clusters. From
the DES Y1 data covering a sky area of $\sim$1800 deg$^2$, 6729
redMaPPer clusters in the redshift range of $0.2<z\lesssim 0.8$ were
identified \citep{rrh+16}. Cross-matching with our clusters in the
same sky region and the redshift range shows that 95\% of redMaPPer
clusters can be detected within a redshift difference of $0.05(1+z)$
and a projected distance of 1.5\,$r_{500}$ from our clusters. As shown
in Figure~\ref{comp}, the detection fraction depends on richness, and
near 100\% clusters with a richness of $\lambda>50$ are
detected. Understandably some clusters with a small richness are
missed due to the uncertainties of photometric redshifts of galaxies
and hence the member recognition.

\citet[][WH22]{wh22} applied the same algorithm as in this paper to the
DES data and identified 151,244 clusters at $0.1<z<1.5$. About 94\% of
clusters at $z<1$ can be found in the catalog of this work. The detection fraction
is near 100\% for rich clusters of $\lambda>50$ (see
Figure~\ref{comp}).

Three cluster samples obtained from photometric redshifts of the DESI
Legacy Surveys DR9 have also been cross-matched with our catalog.
\citet[][Y21]{yxh+21} used the friend-of-friend algorithm and
identified 5.8 million groups in the redshift range of $z<1$, which
have at least three
members\footnote{https://gax.sjtu.edu.cn/data/DESI.html}.
\citet{zsx+22} applied ``the clustering by Fast Search and Find of
Density Peaks (CFSFDP)'' algorithm and identified 532,810
clusters\footnote{https://www.doi.org/
10.11922/sciencedb.o00069.00003} also in the redshift range of $z<1$.
\citet{ynd+23} presented ``the Clusters from Masses and Photometric
Redshifts (CluMPR)'' algorithm and found 309,115
clusters\footnote{https://zenodo.org/record/8157752} in the redshift
range of $0.1<z<1$. The Y21 catalog contains many galaxy groups down
to a very low mass. We here choose the 444,149 clusters with the halo
masses ($M_h$) above our threshold after the mass conversion function
$M_{500}=0.55\,M_h$ \citep{hk03} is applied. In our cluster catalog,
we detect 92\% of Y21 clusters, 72\% of CFSFDP clusters and 86\% of
CluMPR clusters, respectively. The detection fractions vary as a
function of cluster mass or richness.

Noticed that some clusters in previous catalogs are not
detected by our algorithm. We investigate and find three reasons. The
first is that a cluster is located in a cluster-binary system,
probably just starting the merging process. We find both, but their
small separation ($<1.5\,r_{500}$) leads us to regard them as one
cluster. The poorer one is removed during our cleaning stage but are
listed in the other catalogs with a different cluster center. Second,
the unmatched clusters in previous catalogs have no BCG-like
galaxies. The rest are clusters with BCG-like galaxies, but their
richness and the total number of member galaxy candidates are below
our thresholds.

\begin{figure}
  \centering \includegraphics[width = 0.95\columnwidth]{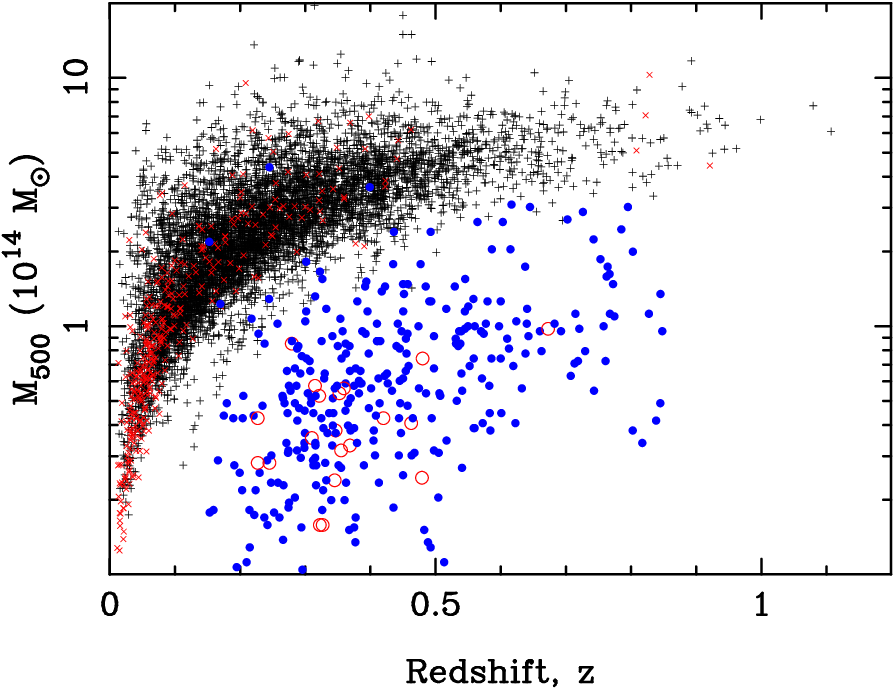}
  \caption{Mass--redshift diagram for clusters in RASS-MCMF
    \citep{khm+23} and eROSITA catalogs \citep{blp+22}. The matched
    RASS-MCMF clusters are shown by plus and the unmatched ones by red
    crosses. The matched eROSITA clusters are shown by blue dots and
    the unmatched ones by red circles.}
\label{compxray}
\end{figure}

\subsection{X-ray clusters}

The X-ray emission of galaxy clusters comes from the hot intracluster
gas. Both the hot gas and galaxy components trace the mass
distribution inside a cluster. In principle, any optical cluster
should have a counterpart in X-ray. We cross-match our clusters from
the DESI Legacy Surveys with the X-ray clusters identified from {\it
  ROSAT} All-Sky Survey \citep{khm+23} and eROSITA
\citep{lbg+22,blp+22}. We find that most {\it ROSAT} clusters are
massive, while the eROSITA clusters are much less massive, as shown in
Figure~\ref{compxray}.

\citet{khm+23} presented a large X-ray cluster sample based on joint
analysis of the second RASS source catalog (2RXS) and the DESI Legacy
Surveys data. They searched for optical counterparts of 2RXS sources
by using a multi-component matched filter (MCMF) algorithm
\citep{kmd+18}. The redshift, optical richness, and X-ray luminosity
are derived, and a probability of being a random superposition $f_{\rm
  cont}$ is given. The RASS-MCMF catalog contains 8449 X-ray clusters
at $z<1$ with a purity of 90\%.  In the sky coverage of this work,
there are 7898 RASS-MCMF clusters. We find that 90\% of them are
detected in this paper. The matched and unmatched RASS-MCMF clusters
are shown in the mass--redshift diagram in Figure~\ref{compxray}. The
detection fractions increase to 93\% and 95\% for the RASS-MCMF
subsamples with the purities of 95\% and 99\%, respectively.

The eROSITA all-sky survey is 25 times more sensitive than the RASS,
and it aims to detect about 50,000--100,000 galaxy clusters
\citep{mpb+12}. The data release of the eROSITA Final Equatorial-Depth
Survey (eFEDS) covers an area of $\sim$ 140 deg$^2$. \citet{lbg+22}
detected 542 candidates of galaxy clusters and groups at $0.01<z<1.3$
with an extended X-ray morphology. The sample has the completeness of
40\% and the purity of 80\% down to the flux limit of $10^{-14}$
erg~s$^{-1}$~cm$^{-2}$. We take a subsample of 262 clusters and groups
with an extent likelihood $\ge15$, that has a purity of 90\%. We find
that 91\% eROSITA clusters and groups are detected in this work. The
X-ray candidates of galaxy clusters and groups with a smaller extent
likelihood probably may easily be misclassified from point
sources. Another eFEDS cluster catalog contains 346 galaxy clusters
and groups in the redshift range of $0.1<z<1.3$ with a mass of
$>10^{13}~M_\odot$ \citep{blp+22}, of which 92\% clusters and groups
can be found in our sample (see Figure~\ref{compxray}). After our
paper has been submitted, the first eROSITA All-Sky Survey (eRASS1)
published the largest X-ray cluster catalog, which contains 12,247
clusters in the western Galactic hemisphere \citep{blk+24}, with a
purity of 86\%. About 82\% eRASS1 clusters have been detected in our
sample. The cross-matching indicates that our optical sample has a
good completeness, even for low mass clusters of about $5\times
10^{13}~M_\odot$.

We check the data and also visually inspect the color images of the
DESI Legacy Surveys to investigate why some X-ray clusters are missing
in our cluster catalog. We find that some of them are merging
clusters, with a projected distance between the optical and X-ray
positions larger than the matching radius. Otherwise, the BCGs in some
clusters are not found due to the cut-off caused by photometry
flags. For most of the low-mass clusters, we can see the overdensity
of member galaxy candidates but the BCG stellar mass or the
richness estimated from member galaxy candidates is lower than our
thresholds.

\subsection{SZ clusters}

The SZ effect is the result of inverse Compton scattering of cosmic
microwave background by the hot intracluser gas. The surface
brightness of the SZ effect is independent of redshift, so that
detecting massive clusters via the SZ effect has the advantage to get
clusters at high redshifts \citep{chr02}.

The {\it Planck} satellite carried out an all-sky millimeter survey in
the millimeter bands \citep{planck14}. The second release of the {\it
  Planck} SZ catalog contains 1653 massive clusters with
$M_{500}>2\times10^{14}~M_\odot$ \citep{planck16b}, of which 789
clusters are located in the DESI Legacy Surveys footprint. We find 746
matched clusters in a redshift difference of $0.05(1+z)$, giving the
detection rate of at least 99\%. A careful check shows that another 35
Planck clusters have a poor redshift estimate and can be matched
within a larger redshift difference \citep{wh22}. One undetected
cluster has the total number of galaxy candidates of $N_{\rm gal}=5$,
below our threshold. We inspect the color images of the other seven
clusters, and find no BCG-like galaxies and no overdensity of member
galaxy candidates in the given redshift slice. They are probably false
detections of the SZ clusters.

The ACT has detected galaxy clusters at the frequencies of 148 GHz, 218 GHz
and 227 GHz with an angular resolution of 1.4 arcmin at 148 GHz
\citep{saa+11}. The latest ACT SZ catalog includes 4195 massive
clusters of $M_{500}>1.5\times10^{14}~M_\odot$ at redshifts
$0.04<z<1.95$ from a sky area of 13,211 deg$^2$ \citep{hsn+21}, of
which 3914 clusters are located in the DESI Legacy Surveys sky area.
The SPT is another excellent facility for detecting SZ clusters at the
frequencies of 95, 150 and 220 GHz \citep{bsd+15}. The SPT cluster
catalog contains 1442 massive clusters up to redshift $\sim1.6$ from
three surveys with different regions and depths
\citep{bsd+15,bbs+20,hbs+20,bka+23}.
We detect 95\% ACT clusters and 93\% SPT clusters in our optical
cluster sample. The other non-detected clusters are mostly at high
redshifts of $z>1$.

\begin{figure}
  \centering \includegraphics[width = 0.9\columnwidth]{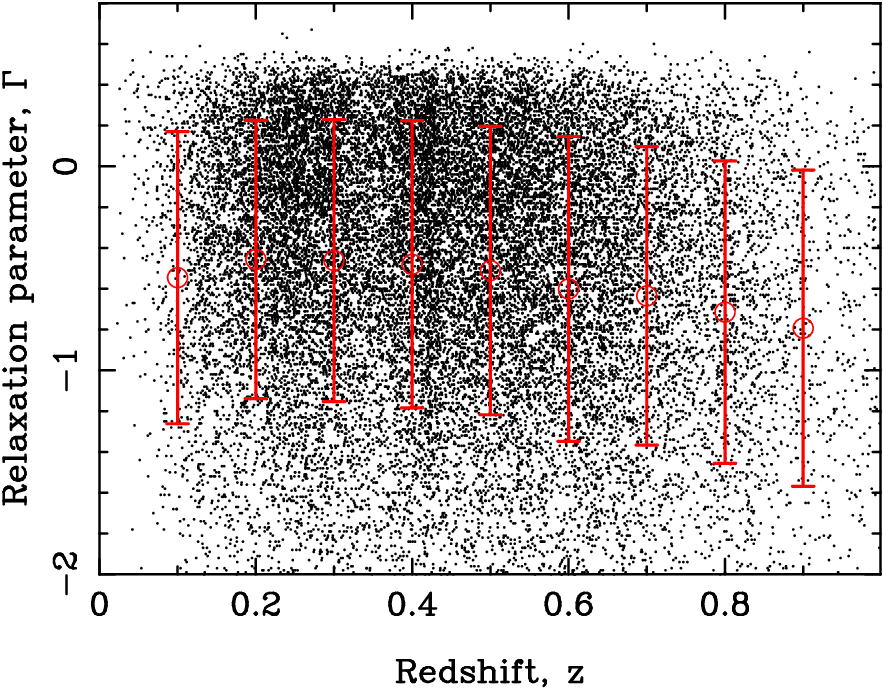}
  \centering \includegraphics[width = 0.9\columnwidth]{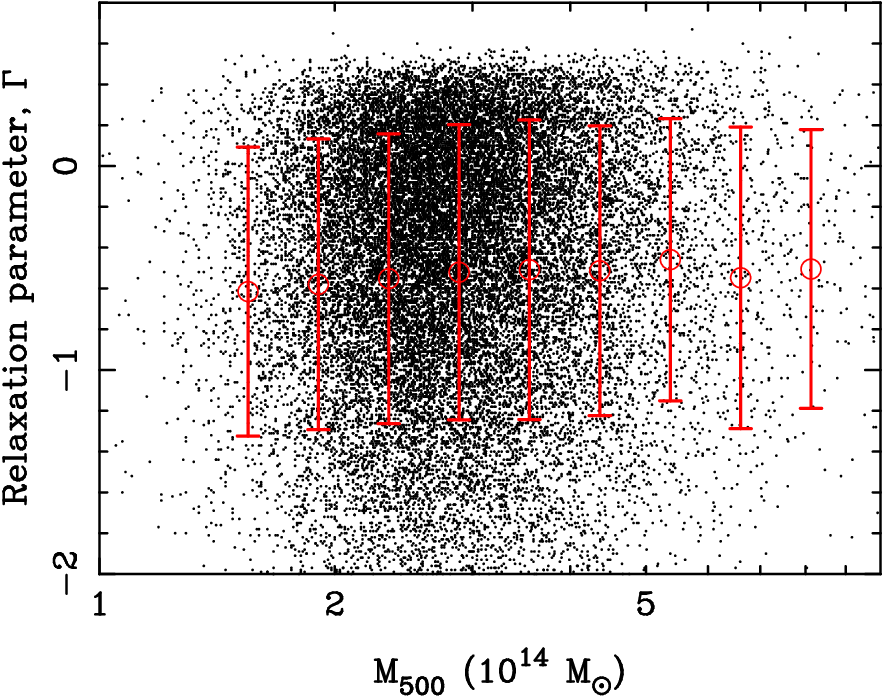}
\caption{The distribution of relaxation parameter $\Gamma$ for 28,038
  massive clusters, indicating no significant evolution with redshift
  (upper panel)and non-dependence with cluster mass (lower panel). The
  open circles and the error bars represent the mean value and scatter
  of dataset, respectively.}
\label{dyna}
\end{figure}

\section{Evolution of dynamical states of clusters and BCGs}

Following \citet{wh13}, we assess the dynamical states of rich
clusters by using the cluster substructures shown by member galaxy
distribution. In addition, we discuss the growth of BCG stellar
masses with redshift.

\subsection{Dynamical states of clusters}
\label{dynamics}

Dynamical states are fundamental for many studies of galaxy clusters
because they contain the information on the assembly history of
clusters \citep{st08}. The dynamical parameters affect the mass
determination of clusters in cosmological studies \citep{see+03}, and
constrain astrophysics processes inside clusters, e.g. gas cooling,
heating, and electron acceleration \citep{bhg+08,fgg+12}. Clusters
have various dynamic states, and are often classified simply as
relaxed clusters and unrelaxed clusters. About 30\%--80\% clusters
show significant substructures in optical or X-ray, and they are in
unrelaxed states \citep[e.g.][]{ds88,srt+08,ddy+21}. It has been
suggested that dynamical states and substructures significantly evolve
with redshift \citep[e.g.][]{mjf+08,me12}, but not conclusive
\citep[e.g.][]{wbc13,wh13,gbb+22}, mainly due to the criteria for
relaxed clusters and the bias of selected sample \citep{emp11}.

\citet{wh13} presented a method to measure dynamical states of
galaxy clusters based on optical photometric data. First, a
smoothed optical map is obtained from the 2-dimensional distribution
of member galaxies weighting their luminosities. Then, three
quantities are calculated: the asymmetry, the ridge flatness, and the
normalized deviation of the smoothed map within the region of
$r_{500}$. The asymmetry measures the asymmetry of the distribution of
smoothed optical light around the cluster center. The ridge flatness
measures the relative steepness of the ridge direction with the most
flat light profile to the other directions.  The normalized deviation
measures the deviation of the optical map to a model. After testing a
cluster sample, three measurements are combined to define a relaxation
parameter, $\Gamma$, which has been used to quantify the dynamical
states of a sample of 2092 rich clusters. This relaxation parameter is
well correlated with the dynamical parameters obtained from X-ray
measurements \citep{wh13,yh20}.

For 28,038 rich clusters in our catalog which have more than 30
member galaxy candidates, we calculate their $\Gamma$ values as listed
in Table~\ref{tab2}.  This is the largest sample of optical clusters
with dynamical state quantified. The distribution of $\Gamma$ values
is shown in Figure~\ref{dyna}. Dynamical states of clusters have a
wide distribution of $\Gamma$ values, not a bimodal distribution for
relaxed and unrelaxed states. No significant variation of
$\Gamma$ with redshift or cluster mass is found, consistent with the
conclusions we previously obtained from the SDSS clusters \citep{wh13}
and \citet{nmb+17} obtained from a X-ray cluster sample.

If relaxed and unrelaxed clusters have to be divided literally by
separating them at $\Gamma=0$, one can get 26.5\% of rich clusters
identified in this paper being dynamically relaxed, which is
consistent with 28\% for the rich SDSS clusters by \citet{wh13}.

\begin{figure}
\centering \includegraphics[width = 0.9\columnwidth]{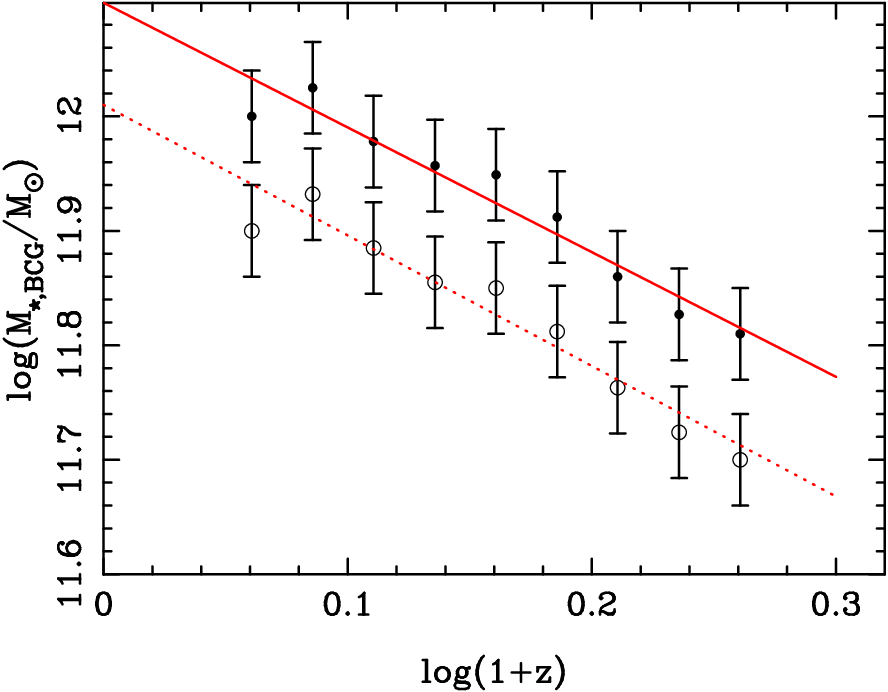}
\caption{To keep a constant comoving number density of BCGs, the
  stellar mass threshold of BCGs as a function of redshift. The dots
  and open circles represent the thresholds starting from a
  reference value of $M_{\star,\rm BCG}=10^{12}~M_\odot$ and
  $10^{11.9}~M_\odot$ at $z=0.1-0.2$, respectively. The solid and
  dashed lines are the best fits the data in dots and open circles,
  respectively.}
\label{BCGevo}
\end{figure}

\begin{figure}
\centering \includegraphics[width = 0.9\columnwidth]{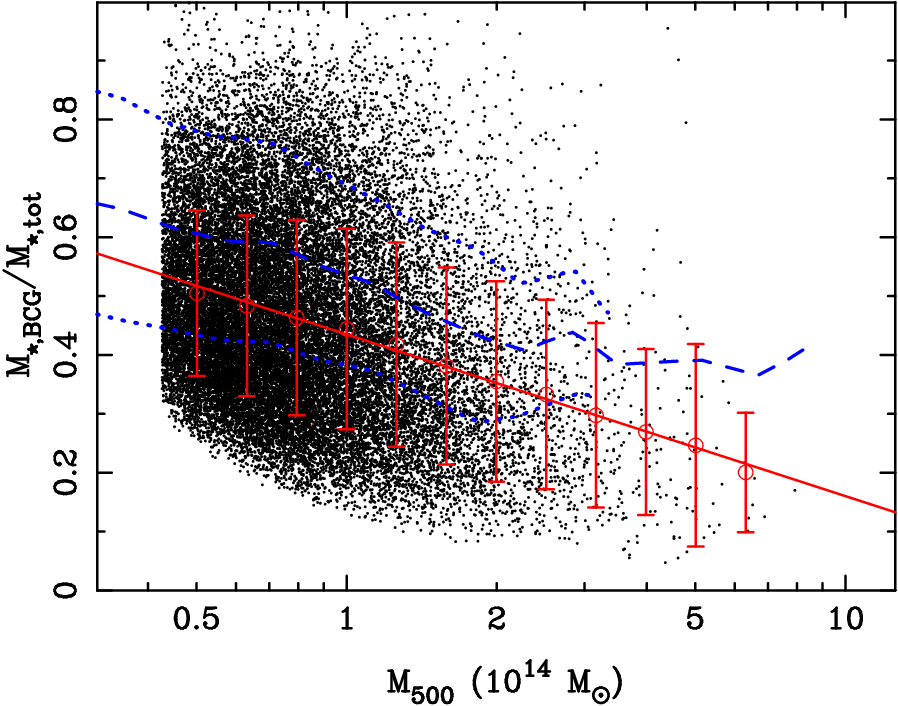}
\caption{The fraction for BCG stellar mass of total stellar mass of
  galaxies in clusters, plotted as a function of cluster mass. The
  average and data dispersions are given by circles and error-bar,
  with a fitted line. The dashed line and dotted line come from the
  simulations by \citep{kvm+18} for the mean fraction and the
  dispersion.}
\label{BCGfra}
\end{figure}

\subsection{Growth of BCGs}

The BCGs have intriguing properties, and their evolution is tightly
related to the host clusters \citep{vbk+07,whl12,yw22}. The stellar
mass of BCGs can be a good tracer to understand their evolution.  From
simulations by \citet{db07}, the BCG stellar mass grows by a factor of
3 from $z=1$ to $z=0$. Diverse results were later obtained by
observations \citep[e.g.][]{lbg+13, zmm+16, wh21}. The infrared
luminosities were found to be consistent with a passive evolution
model since $z\sim1$, suggesting no significant change in the BCG stellar masses
\citep{wad+08,scs+10}. \citet{lsm+12} corrected the relation between BCG
stellar mass and cluster mass, and found that the stellar mass of BCGs
increases by a factor of $1.8\pm0.3$ from $z=0.9$ to 0.2.

Here, we investigate the growth of the BCG stellar mass by a new
approach. Assuming that the merger between BCGs is negligible, the
comoving number density of BCGs should be a constant over cosmic time
for a volume-limited complete sample. Since all BCGs in our large
cluster sample have a mass $M_{\star,\rm BCG}>10^{11}~M_\odot$, we can
take a high threshold of BCG stellar mass for clusters in the redshift
range of $z=0.1-0.2$ as the reference, and then reduce the threshold
of BCG stellar mass to keep the same comoving density of BCGs in other
redshift bins, as shown by Figure~\ref{BCGevo}. For BCGs with a
threshold of BCG stellar mass $M_{\star,\rm BCG}>10^{12}~M_\odot$ in
the redshift range of $z=0.1-0.2$, the thresholds in other redshift
bins can be fitted by a linear law in the log-log space,
\begin{equation}
\log (M_{\star,\rm BCG})=(12.10\pm0.02)-(1.09\pm0.10)\log (1+z).
\label{growth_1}
\end{equation}
This relation implies that the stellar mass grows by a factor of
$2.12^{+0.16}_{-0.14}$ since $z=1$. If a threshold of $M_{\star, \rm
  BCG}=10^{11.9}~M_\odot$ is chosen, one can get fit to the
thresholds by
\begin{equation}
\log (M_{\star,\rm BCG})=(12.01\pm0.02)-(1.14\pm0.11)\log (1+z),
\label{growth_2}
\end{equation}
which means that the stellar masses of BCGs grow by a factor of
$2.20^{+0.17}_{-0.16}$ since $z=1$. Our results are consistent with
previous works based on observation data \citep{lsm+12,zmm+16,wh21}.

In optical, the BCG is usually dominant inside a cluster. The
dominance of BCGs in clusters can be expressed by the fraction of BCG
stellar mass relative to the total stellar mass (including the BCG) if
the members of clusters are highly complete. We therefore take all
clusters at $z<0.5$ for this work. As seen in Figure~\ref{BCGfra}, the
fraction of BCG stellar mass decreases with cluster mass, indicating
the BCGs are less dominant in more massive clusters with more member
galaxies, consistent with the results from simulations
\citep{kvm+18}. The mean fraction depends on cluster mass by the
relation
\begin{equation}
\frac{M_{\star,\rm BCG}}{M_{\star,\rm tot}}=(0.435\pm0.003)-(0.275\pm0.007)\log M_{500}.
\end{equation}

In addition, we investigate how many BCG-like galaxies are located in
the clusters. BCG-like galaxies are considered as being cluster
members if they have a projected separation within 1.5\,$r_{500}$ and
a redshift difference less than $\Delta_z$. We find that about 76\% of
massive galaxies with a mass of $M_\star \sim 10^{12}~M_{\odot}$, or
about 50\% of galaxies of $M_\star \sim 10^{11.7}~M_{\odot}$ or 10\%
at $M_\star \sim 10^{11}~M_{\odot}$, are located in the identified
clusters. Therefore, finding out massive BCG-like galaxies is a useful
first step to identifying galaxy clusters.

\section{Summary}

By using photometric redshifts and available spectroscopic redshifts
of galaxies in the joint DESI Legacy Surveys and WISE data, we obtain
a large catalog of 1.58 million galaxy clusters and investigate their
properties.

We first estimate the photometric redshifts for 295.6 million galaxies
in the latest released data in the $griz{\rm W1}{\rm W2}$ bands by
using the nearest-neighbor algorithm. The redshift uncertainty is
about 0.013 for bright galaxies of $z_{\rm m}<21$ and increases to
0.053 for faint galaxies of $z_{\rm m}>22$. The stellar mass is
estimated from the $z{\rm W1}$-band luminosities for galaxies in the
DR10 and also those galaxies in the DR9 if their photometric redshifts
are estimated.

Then, we identify a large sample of 1.58 million galaxy clusters based
on the overdensity of total galaxy stellar mass within a redshift
slice around BCG-like galaxies. Among them, 877,806 clusters are
identified for the first time. The redshift distribution extends to
$z<1.5$, and 338,841 clusters have spectroscopic redshift. The
uncertainty of cluster photometric redshift is about 0.010. The cluster
richness has a good correlation with cluster mass, which indicates
that the identified clusters
have an equivalent mass of $M_{500}\ge 0.47\times10^{14}~M_{\odot}$. Our
cluster catalog includes most of the massive clusters in previous optical,
X-ray and SZ cluster catalogs.

We assess the dynamical states for 28,038 rich clusters based
on the substructures of the distribution of cluster member
galaxies. No significant evolution of dynamical states is shown.
In addition, we find that the stellar mass of BCGs increases by a factor
of 2 after $z=1$. The fraction for the BCG stellar mass relative to
the total stellar mass decreases with cluster mass, implying less
dominance of BCGs in richer clusters.

\section*{acknowledgments}

We thank the referee for the valuable comments that helped to improve
the paper. The authors are partially supported by the National Natural
Science Foundation of China (Grant Numbers 11988101, 11833009 and
12073036), the Key Research Program of the Chinese Academy of Sciences
(Grant Number QYZDJ-SSW-SLH021). We also acknowledge the support of
the science research grants from the China Manned Space Project with
Numbers CMS-CSST-2021-A01 and CMS-CSST-2021-B01.

The Legacy Surveys consist of three individual and complementary
projects: the Dark Energy Camera Legacy Survey (DECaLS; Proposal ID \#
2014B-0404; PIs: David Schlegel and Arjun Dey), the Beijing-Arizona
Sky Survey (BASS; NOAO Prop. ID \# 2015A-0801; PIs: Zhou Xu and
Xiaohui Fan), and the Mayall z-band Legacy Survey (MzLS; Prop. ID \#
2016A-0453; PI: Arjun Dey). DECaLS, BASS, and MzLS together include
data obtained, respectively, at the Blanco telescope, Cerro Tololo
Inter-American Observatory, NSF’s NOIRLab; the Bok telescope, Steward
Observatory, University of Arizona; and the Mayall telescope, Kitt
Peak National Observatory, NOIRLab. Pipeline processing and analyses
of the data were supported by NOIRLab and the Lawrence Berkeley
National Laboratory (LBNL). The Legacy Surveys project is honored to
be permitted to conduct astronomical research on Iolkam Du’ag (Kitt
Peak), a mountain with particular significance to the Tohono O’odham
Nation.

NOIRLab is operated by the Association of Universities for Research in
Astronomy (AURA) under a cooperative agreement with the National
Science Foundation. LBNL is managed by the Regents of the University
of California under contract to the U.S. Department of Energy.

This project used data obtained with the Dark Energy Camera (DECam),
which was constructed by the Dark Energy Survey (DES)
collaboration. Funding for the DES Projects has been provided by the
U.S. Department of Energy, the U.S. National Science Foundation, the
Ministry of Science and Education of Spain, the Science and Technology
Facilities Council of the United Kingdom, the Higher Education Funding
Council for England, the National Center for Supercomputing
Applications at the University of Illinois at Urbana-Champaign, the
Kavli Institute of Cosmological Physics at the University of Chicago,
Center for Cosmology and Astro-Particle Physics at the Ohio State
University, the Mitchell Institute for Fundamental Physics and
Astronomy at Texas A\&M University, Financiadora de Estudos e
Projetos, Fundacao Carlos Chagas Filho de Amparo, Financiadora de
Estudos e Projetos, Fundacao Carlos Chagas Filho de Amparo a Pesquisa
do Estado do Rio de Janeiro, Conselho Nacional de Desenvolvimento
Cientifico e Tecnologico and the Ministerio da Ciencia, Tecnologia e
Inovacao, the Deutsche Forschungsgemeinschaft and the Collaborating
Institutions in the Dark Energy Survey. The Collaborating Institutions
are Argonne National Laboratory, the University of California at Santa
Cruz, the University of Cambridge, Centro de Investigaciones
Energeticas, Medioambientales y Tecnologicas-Madrid, the University of
Chicago, University College London, the DES-Brazil Consortium, the
University of Edinburgh, the Eidgenossische Technische Hochschule
(ETH) Zurich, Fermi National Accelerator Laboratory, the University of
Illinois at Urbana-Champaign, the Institut de Ciencies de l’Espai
(IEEC/CSIC), the Institut de Fisica d’Altes Energies, Lawrence
Berkeley National Laboratory, the Ludwig Maximilians Universitat
Munchen and the associated Excellence Cluster Universe, the University
of Michigan, NSF’s NOIRLab, the University of Nottingham, the Ohio
State University, the University of Pennsylvania, the University of
Portsmouth, SLAC National Accelerator Laboratory, Stanford University,
the University of Sussex, and Texas A\&M University.

BASS is a key project of the Telescope Access Program (TAP), which has
been funded by the National Astronomical Observatories of China, the
Chinese Academy of Sciences (the Strategic Priority Research Program
“The Emergence of Cosmological Structures” Grant \# XDB09000000), and
the Special Fund for Astronomy from the Ministry of Finance. The BASS
is also supported by the External Cooperation Program of the Chinese
Academy of Sciences (Grant \# 114A11KYSB20160057), and Chinese
National Natural Science Foundation (Grant \# 12120101003, \#
11433005). The Legacy Survey team makes use of data products from the
Near-Earth Object Wide-field Infrared Survey Explorer (NEOWISE), which
is a project of the Jet Propulsion Laboratory/California Institute of
Technology. NEOWISE is funded by the National Aeronautics and Space
Administration.

The Legacy Surveys imaging of the DESI footprint is supported by the
Director, Office of Science, Office of High Energy Physics of the
U.S. Department of Energy under Contract No. DE-AC02-05CH1123, by the
National Energy Research Scientific Computing Center, a DOE Office of
Science User Facility under the same contract; and by the
U.S. National Science Foundation, Division of Astronomical Sciences
under Contract No. AST-0950945 to NOAO.

The Photometric Redshifts for the Legacy Surveys (PRLS) catalog used
in this paper was produced, thanks to funding from the U.S. Department
of Energy Office of Science, Office of High Energy Physics via grant
DE-SC0007914.

\bibliographystyle{aasjournal}
\bibliography{desi}

\end{document}